\documentclass[superscriptaddress,groupedaddress,nofootnoteinbib,11pt]{article}
\pdfoutput=1
\usepackage{jcappub}
\usepackage{bm}

\usepackage{graphicx}

\usepackage{amsfonts}
\usepackage{amssymb}
\usepackage{xcolor}

\definecolor{mine}{rgb}{0.2,0.1,0.7}
\definecolor{bb}{rgb}{0.3, 0.5, 1}
\definecolor{bg}{rgb}{0.1, 0.1, 0.5}

\newcommand{\re}{{\rm re}}
\newcommand\spart{\;\raise1.0pt\hbox{/}\hskip-6pt\partial}
\newcommand\spartb{\;\overline{\raise1.0pt\hbox{/}\hskip-6pt\partial}}
\newcommand{\vertsp}{\vphantom{\displaystyle{\dot a \over a}}}
\newcommand{\ve}{{(1)}}
\newcommand{\te}{{(2)}}
\newcommand{\F}{F}
\newcommand{\n}{{\hat n}}
\newcommand{\Y}{y}
\newcommand{\calD}{{\cal D}}
\newcommand{\calP}{{\cal P}}
\newcommand{\refeq}[1]{(\ref{#1})}

\newcommand{\gr}[1]{\mathbf{#1}}
\newcommand{\ii}{\mathrm{i}}

\newcommand{\parq}{{\frac{\partial}{\partial \ln q}}}
\newcommand{\parqtwo}{{\frac{\partial^2}{\partial \ln q^2}}}

\newcommand{\dd}{{\rm d}}
\newcommand{\BBdis}{{I_{\rm BB}}}

\newcommand*{\msk}{\\[0.25cm]} 

\begin{document}

\title{Spectral distortions in the cosmic microwave background polarization}

\author[1,2]{S\'ebastien Renaux-Petel,}
\author[3]{Christian Fidler,}
\author[2,4]{Cyril Pitrou,}
\author[5]{and Guido W. Pettinari}

\affiliation[1]{Laboratoire de Physique Th\'eorique et Hautes
   Energies, Universit\'e  Pierre \& Marie Curie - Paris VI, CNRS-UMR 7589, 4 place Jussieu, 75252 Paris, France}
\affiliation[2]{Sorbonne Universit\'es, Institut Lagrange de Paris,
  98 bis Bd Arago, 75014 Paris, France}
\affiliation[3]{Institute of Cosmology and Gravitation,
Dennis Sciama Building, Burnaby Road, Portsmouth, PO1 3FX (United
Kingdom)}   
\affiliation[4]{Institut d'Astrophysique de Paris,
Universit\'e Pierre~\&~Marie Curie - Paris VI,
CNRS-UMR 7095, 98 bis Bd Arago, 75014 Paris, France}
\affiliation[5]{Department of Physics \& Astronomy, University of Sussex,
Brighton BN1 9QH, UK}



\abstract{We compute the spectral distortions of the Cosmic Microwave
  Background (CMB) polarization induced by non-linear effects in the
  Compton interactions between CMB photons and the flow of intergalactic
  electrons. This signal is of the $y$-type and is dominated by contributions arising from
  the reionized era. We stress that it is not shadowed by the thermal
  SZ effect which
  has no equivalent for polarization. We decompose its angular
  dependence into $E$- and $B$-modes, and we calculate the
  corresponding power spectra, both exactly and using a suitable
  Limber approximation that allows a simpler numerical evaluation. We
  find that $B$-modes are of the same order of magnitude as $E$-modes. Both spectra are relatively flat, peaking around
  $\ell=280$, and their overall amplitude is directly related to the optical
  depth to reionization. Moreover, we find this effect to be one order of
  magnitude larger than the non-linear kinetic Sunyaev-Zel'dovich effect in galaxy clusters. Finally, we discuss how to improve the detectability of our signal by cross-correlating it with other quantities sourced by the flow of intergalactic electrons.
  }

\maketitle


\section{Introduction}

The Planck mission has recently provided exquisite maps of the temperature anisotropies of the Cosmic Microwave Background (CMB) radiation, from which angular spectra and cosmological parameters were extracted~\cite{Ade:2013zuv}; measurements of the polarized signal, especially in the $E$-modes, will be released soon.
More generally, the CMB has been the leading probe of cosmology in the past decades, mainly because its physics is well described by linear perturbation theory, in sharp contrast with the highly non-linear dynamics of large scale structure formation, which involves density fluctuations of order unity.
However, the precision of current CMB measurements is such that non-linear effects cannot be ignored anymore. For example, lensing effects by the foreground distribution of matter have to be taken into account in the angular power spectra of the CMB~\cite{Lewis:2006fu}, and the lensing potential has even been reconstructed from the connected four-point correlation function of the temperature
fluctuations~\cite{Ade:2013tyw}. Furthermore, the three-point correlation function has been used to
provide stringent limits on primordial non-Gaussianity and to detect
the correlation between the gravitational lensing and the
integrated Sachs-Wolfe effect \cite{Ade:2013ydc}. It thus becomes clear that the CMB science will now be driven by the study of non-linear effects.

In this respect, an effect that has comparatively received much less attention than the ones aforementioned is the study of the frequency dependence of the CMB. At linear level, no deviation from a blackbody spectrum is generated\footnote{In the absence of phenomena that inject energy into the primordial plasma.} and the spectral information is solely characterized by a direction-dependent temperature. However, in full generality, the observed CMB radiation depends non-trivially both on the direction of observation and on the energy, or frequency of the photons received. Future possible CMB experiments, such as PRISM~\cite{Andre:2013afa} or PIXIE~\cite{Kogut:2011xw}, will map the intensity and linear polarization over the full sky in many spectral channels, allowing to probe with great accuracy the deviations from a pure blackbody spectrum, or so-called spectral distortions. Besides the well-known and already observed thermal Sunyaev-Zel'dovich (tSZ) effect \cite{SZoriginal}, there exists a guaranteed
minimal signal, generated before recombination, within the reach of these experiments \cite{Sunyaev:2013aoa}. More generally, spectral distortions are a powerful probe of many physical phenomena that inject energy into the primordial plasma, such as dark matter annihilation or Silk damping of primordial density perturbations (see e.g. Refs.~\cite{Sunyaev:2013aoa,Chluba:2011hw,Chluba:2012gq}). 

At the non-linear level, Compton collisions also induce deviations from a pure blackbody
spectrum~\cite{Stebbins2007} from the non-linear couplings of photons with the
bulk velocity of baryons, $v_{\rm b}$, during the reionized era. These distortions are mainly of
the $y$-type and their angular spectrum has been computed
numerically~\cite{Pitrouysky}. However, since distortions of this type
are also generated by the tSZ effect,
this signal is shadowed by the contributions coming from unresolved
point sources~\cite{Ade:2013qta}. Indeed, the thermal energy of electrons in a typical galaxy cluster, $k_{\rm B}
T_{\rm e}$, is four orders of magnitude larger than their bulk motion kinetic energy, $m_{\rm e} v_{\rm b}^2/2$.

The situation is different when considering the frequency dependence
of the CMB polarization, which is the topic of this paper. In this
case, the tSZ effect is subdominant since it introduces only a correction
of order $ 6 k_{\rm B}T_{\rm e}/(m_{\rm e} c^2) \sim 10^{-1}$ to the leading
effect~(see Eq.~(36) of Ref.~\cite{Challinor:1999yz}), the latter
being due to the local quadrupole of the radiation in the baryon rest frame
\cite{Hu:1999vq,Baumann:2002es,Lavaux:2003qf} (note that there is
also a contribution to polarization when considering
effects at second order in optical depth with directional dependence
like inside clusters, see \S~3.3 of Ref.~\cite{Sunyaev:1980nv} and
\S~4.2 of Ref.~\cite{Sazonov:1999zp}). This quadrupole can be intrinsic to the CMB as it
was sourced by free-streaming since the last scattering surface, or it can be due to
the local motion of baryons. However, only the second physical process, known in this
context as the non-linear kinetic Sunyaev Zel'dovich  (kSZ) effect~\cite{Sunyaev:1980nv}, has a spectral
dependence of the $y$-type~\cite{Sazonov:1999zp,Challinor:1999yz}.

In this article, we consider the non-linear kSZ effect due to the bulk velocities of baryons in the intergalatic medium, and compute its unavoidable contribution to the $y$-type spectral distortion in polarization. We compare to previous results for the non-linear kSZ effect
generated by clusters \cite{Baumann:2002es}, and find that the intergalatic contribution is one order of
magnitude larger.\\

The outline of the paper is as follows. In \S~\ref{PartDescription}, we first present the description of
spectral distortions of the $y$-type in the linearly polarized CMB. We
then discuss the dynamics of such type of distortion from the
Boltzmann equation, and derive the general solution for the angular multipoles of the $E$- and $B$-modes of distortions.
In \S~\ref{Power-spectra}, we obtain the exact expression for their angular
power spectra, and present a suitable Limber approximation. We finally
present numerical results in \S~\ref{SectionNumResults}, both with the
exact and approximate methods, and discuss our results. We conclude in \S~\ref{sec:conclusion} and gather some technical details in four appendices.

\section{Describing the spectral distortions of polarization}\label{PartDescription}

\subsection{The distribution function and its spectral decomposition}

In this section, we introduce our set-up and notations, following Ref.~\cite{Naruko2013} to which we refer the reader for more details. The description of polarized radiation is formulated by introducing a field of tetrad basis, \textit{i.e.} a set of four vector fields $e_{(a)}^{\mu}$ $(a=0,1,2,3)$ that satisfy 
\begin{equation}
\eta_{(a)(b)}=g_{\mu\nu}\,e_{(a)}{}^{\mu} e_{(b)}{}^{\nu}, \quad\quad g_{\mu \nu}=\eta_{(a)(b)} e^{(a)}{}_{\mu} e^{(b)}{}_{\nu}
\end{equation}
where $g_{\mu \nu}$ is the spacetime metric and $\eta_{(a)(b)}$ is the
Minkowski metric. This defines
everywhere a local frame which is used to separate the magnitude of the photon momentum from its direction~\cite{Pitrou2008}. We project the photon momentum $p^{\mu}$ onto the set of tetrads, using $p^{\mu}=p^{(a)} e_{(a)}{}^{\mu}$, and further define the comoving momentum $q$ and the photon direction $n^{(i)}$ as 
\begin{equation}
p^{(0)}=\frac{q}{a}\,, \quad\quad p^{(i)}= \frac{q}{a} n^{(i)} \quad (i=1,2,3)
\end{equation}
where $n_{(i)} n^{(i)}=1$ and $a$ is the cosmological scale
factor.
Physically, with the choice $\boldsymbol{e}^{(0)} \propto {\rm
  d} \eta $ where $\eta$ denotes conformal time, the above $q$ is
simply proportional to the energy measured by an observer whose wordline is orthogonal
to constant time hypersurfaces: $q=a E_{{\rm phys}}$. 

To describe the polarized radiation in the statistical description, one has
to introduce a Hermitian tensor-valued distribution function,
$\,f_{\mu\nu}(\eta,\gr{x},p^{(i)})\,$, such that
\begin{equation}
  \epsilon^\mu\;\epsilon^{*\nu}\;f_{\mu\nu}(\eta,\gr{x},p^{(i)})\;
\end{equation}
is the number density in phase space of photons at
($\eta$,\,$\gr{x}$,\,$p^{(i)}$) with polarization state vector $\epsilon^\mu$ (see Refs.~\cite{Pitrou2008, beneke:2010a} and references therein).  We also
define the direction 4-vector of photons $n^{\mu}$ by $n^{\mu} \equiv  n^{(i)} e_{(i)}{}^{\mu} $, with which we define the projection operator, also often called the screen projector, as
\begin{equation}
 S_{\mu \nu}
 \equiv g_{\mu \nu} + e^{(0)}{}_\mu e^{(0)}{}_\nu - n_\mu n_\nu \,.
\end{equation}
By definition, $S_{\mu \nu}$ is a projector onto the two dimensional hypersurface
orthogonal to both $e^{(0)}{}_\mu$ and $n_\mu$, that is, the
plane on which ``lives'' the complex-valued and Hermitian distribution
function $f_{\mu \nu}$ describing the polarization of
radiation\footnote{The distribution function verifies indeed $f_{\mu
    \nu} =S_\mu{}^\alpha S_{\mu}{}^\beta f_{\alpha \beta}$. See
  e.g. Refs.~\cite{Tsagas:2007yx,Naruko2013} for more
  details.}. 
The latter can be decomposed, in the absence of circular polarization - as relevant in the standard cosmological context - as\footnote{Note that the $P_{\mu \nu}$ defined in this paper is equal to $2 P_{\mu \nu}$ of Refs.~\cite{Naruko2013,Tsagas:2007yx}.}
\begin{equation}
 f_{\mu \nu}  \equiv \frac{1}{2} \left(I\:
 S_{\mu \nu} \,+\, P_{\mu \nu} \right)  \,,
 \label{Pol}
\end{equation}
where {\it I} is the intensity and the symmetric traceless tensor $P_{\mu \nu}$ encodes the two degrees of linear polarization, which are parametrized by the $Q$ and $U$ Stokes parameters, see \S~\ref{Power-spectra}.

On the background Friedmann-Lema\^itre-Robertson-Walker (FLRW) space-time, the distribution function is
characterized solely by the intensity $I$, which follows a Planck
distribution whose temperature $\overline T$ depends only on time $\eta$ due to
the symmetries of the FLRW universe. From the background Boltzmann
equation, it is easily found that $\overline T(\eta) = T_0\,a_0/ a(\eta)$
where $a_0$ and $T_0$ are respectively the scale factor and the background
temperature today, so that, on the background, the intensity of the photons is described by a redshifted blackbody distribution,
\begin{equation}
\overline I(\eta,q) \,=\, \BBdis\left(\frac{q}{a(\eta)\overline T(\eta)}\right) \,=\,
\BBdis\left( \frac{q}{a_0 T_0}\right)\,, \quad\quad
 \text{with} \quad \BBdis (x) \equiv \frac{2}{( e^x - 1 )} \,.
\end{equation}

No spectral distortion is generated at first order in perturbation
theory~\cite{dodelson:2003b}, under which approximation the intensity fluctuation can simply be described by a temperature fluctuation independent of the energy $q$. However, Compton scattering does induce spectral distortions at second order \cite{hu:1994a, dodelson:1995a}, which can be described by introducing the so-called $y$ Compton parameter, entering into the Fokker-Planck expansion of the distribution function~\cite{Stebbins2007} as
\begin{align} \label{defy}
 I \Bigl( q,n^{(i)} \Bigr)
 &\;\simeq\; \BBdis \left( \frac{q}{a T(n^{(i)} )} \right)
 \;+\; y \bigl( n^{(i)} \bigr) \, q^{-3} \parq \left[
 q^3 \parq \BBdis \left( \frac{q}{a T(n^{(i)} )} \right) \right] \notag\msk
 &\;=\; \BBdis \left( \frac{q}{a T(n^{(i)} )} \right) \;+\; y \bigl( n^{(i)} \bigr) \,
 \calD_q^2\,\BBdis \left( \frac{q}{aT(n^{(i)} )} \right) \,,
\end{align}
where
\begin{equation}
 \calD_q^2 \;\equiv\; q^{-3} \parq \left( q^3 \parq \right)
 \;=\; \parqtwo\;+\;3 \,\parq \,
\end{equation}
and, for brevity, we have omitted the dependence of all quantities on
the spacetime coordinates $x^\mu$. The only quantities in the expansion that depend on the propagation direction of the photon are the Compton parameter $y$ and the temperature $T$.
This expansion is particularly convenient to describe spectral distortions. Indeed, as the number density of photon is given by $n
\propto a^{-3}\int I\,q^2\,\dd q\,$, the $y$ term does not contribute to
the photon number density and the temperature $T$ is the temperature of the blackbody that would have the same number density. 

Similarly, the spectral dependence of the polarization tensor can be split into a standard component $\calP_{\mu \nu}$, and a spectral distortion component $\Y_{\mu\nu}$ as \cite{Stebbins2007,Naruko2013}
\begin{equation}\label{defyPbase}
 P_{\mu \nu} \Bigl( q, n^{(i)} \Bigr)
 \;\simeq\; - \calP_{\mu\nu} \bigl( n^{(i)} \bigr) \,\parq \,\BBdis
 \left( \frac{q}{a T(n^{(i)} )} \right)
 \;+\; \Y_{\mu \nu} \bigl( n^{(i)} \bigr) \,\calD_q^2 \,\BBdis
 \left( \frac{q}{a T(n^{(i)} )} \right) \,.
\end{equation}
Note that this splitting is slightly different from the case of the
intensity as there is no polarization on the background and hence no
term corresponding to the first one in Eq.~(\ref{defy}). Similarly to
the $y$ parameter, it can be shown that Compton scattering generates a
non-vanishing $\Y_{\mu\nu}$ only beyond first-order perturbation theory
\cite{Naruko2013}. We name this term {\it $y$-type distortion of
  linear polarization} or polarization distortion for short. The component $\calP_{\mu \nu}$,
on the other hand, describes the standard polarization \cite{Seljak:1996gy}. Let us stress eventually that the temperature $T$ entering into Eqs.~\refeq{defy}-\refeq{defyPbase} is a local quantity, which can itself be expanded around the background temperature as
\begin{equation}
 T \bigl(\eta, n^{(i)} \bigr)
 \;\equiv\; \overline T(\eta) \left[ 1 + \Theta \bigl( n^{(i)} \bigr) \right] \,.
\end{equation}

\subsection{Leading order solution of the Boltzmann equation}
\label{LOsolution}

The time evolution of the distribution function $f_{\mu \nu}$ is
governed by the Boltzmann equation. Since the metric
  fluctuations affect the energy of photons in the same proportions
  through free streaming,  that is they affect $\ln q$ and not $q$, and
  given the parameterizations (\ref{defy}) and (\ref{defyPbase}) which are
  based on logarithmic derivatives of $\BBdis$, then the
  Boltzmann equation can be separated into dynamical equations for
  its spectral components $\Theta$ and $y$ for the intensity part, or $\calP_{\mu\nu}$ and
  $\Y_{\mu\nu}$ for the polarized part, which couple only through the collision term. More details of this procedure
can be found in Refs.~\cite{Stebbins2007,Naruko2013,Pitrou:2014ota}. By expanding the
Boltzmann equation at second order in
perturbation theory, one can thus obtain the evolution equation for our
quantity of interest $\Y_{\mu \nu}$. In the tetrad basis, it can be written in the form 
\begin{equation}
\Y_{(i)(j)}' \;+\; n^{(l)} \partial_l \Y_{(i)(j)} \;=\;
\tau'\left(-\Y_{(i)(j)}+C^\Y_{(i)(j)} \right) \;,
\label{Y'}
\end{equation}
where a prime denotes derivation with respect to the conformal time $\eta$. The Thomson interaction rate is given by $\tau' \equiv a \, \overline n_e \,\sigma_{\rm T}$, where $\overline n_e$ is the background density of free electrons and $\sigma_{\rm T}$ is the Thomson scattering cross section.
The explicit and lengthy expression of the collision term $C^\Y_{(i)(j)}$ can be found in Ref.~\cite{Naruko2013}. Fortunately, we do not need its full expression as the leading order term can be easily identified as
\begin{equation}
C^{{\Y}\, {\rm (L.O.)}}_{(i)(j)}\;=\; -\frac{1}{10}\;\left[\,v_{(i)} v_{(j)}\,\right]^{\rm TT}
\label{Cij}
\end{equation}
where $v_{(i)}$ is the difference between the first-order baryon velocity $v_b$ and photon velocity
$v_r$ in the tetrad basis (in the following we shall simply refer to $v_{(i)}$ as the baryon
velocity), and where ${\rm TT}$ denotes the traceless screen-projected part
\begin{equation}
\label{TT}
[v_{(i)} v_{(j)}]^{\rm TT} \;\equiv\; \left[\,S_{(i)}{}^{(k)} S_{(j)}{}^{(l)} - \frac{1}{2} S^{(k) (l)} S_{(i) (j)} \,\right]\; v_{(k)} \,v_{(l)}\,.
\end{equation}
Although the collision term (\ref{Cij}) resembles the
  one derived in Refs.~\cite{Challinor:1999yz,Hu:1999vq,Lavaux:2003qf,Baumann:2002es,Sazonov:1999zp}
for the overall numerical factor and the baryons velocity geometric
contribution $[v_{(i)} v_{(j)}]^{\rm TT}$, its spectral dependence is
different. In our case the spectral dependence is implied by the
definition~(\ref{defyPbase}) and is $\calD_q^2 \,\BBdis$, which corresponds only to spectral
distortions. In these initial references, the spectral
dependence of the collision term is $(\calD_q^2 \,-4 \partial/\partial
\ln q)\BBdis$ and it corresponds to the total collision term,
encompassing both the contributions to the spectral distortion
$y_{\mu\nu}$ and the standard component $\calP_{\mu \nu}$. After
integration over the spectral dependence, this leads to the collision
term for the brightness of the polarized signal (the one presented in Ref.~\cite{Hu:1999vq}),
which still possesses the geometrical dependence $[v_{(i)} v_{(j)}]^{\rm
  TT}$ but then has a different numerical factor as it corresponds to
the collision term of a variable proportional to $\calP_{\mu \nu}+y_{\mu \nu}$
in our framework.

\begin{figure}[!h]
  \center
  \includegraphics[width=0.7\textwidth]{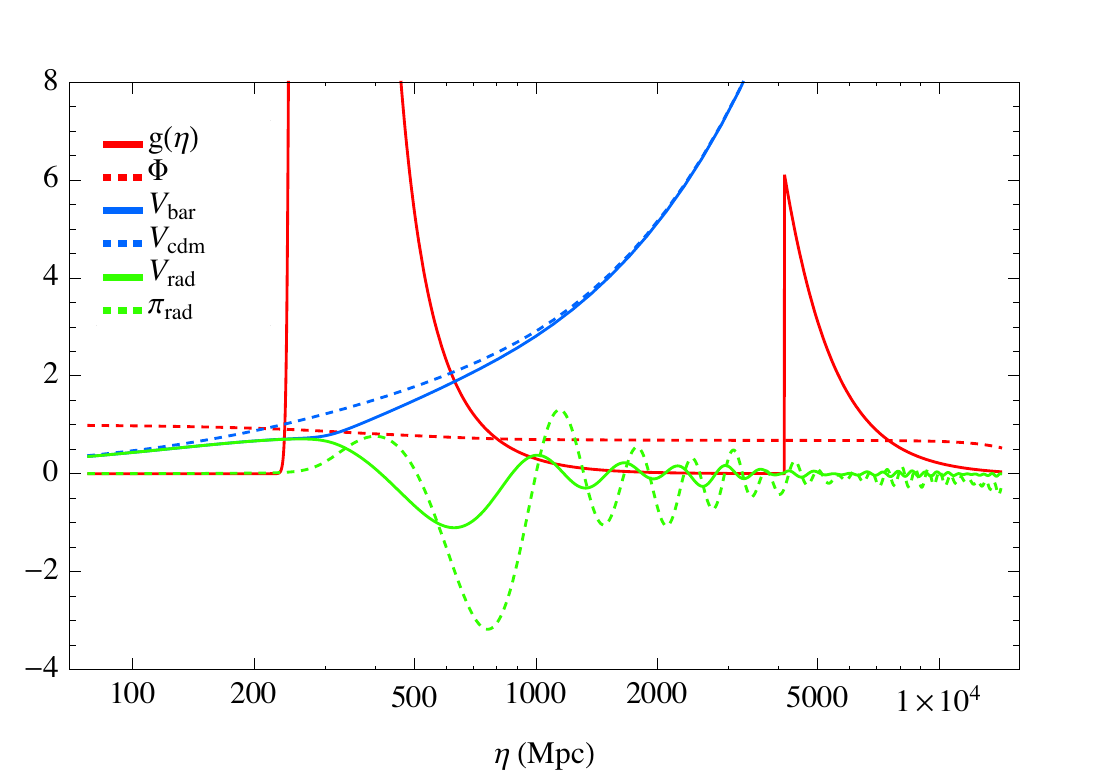}
      \caption{Red continuous line: visibility function \refeq{g} (magnified by a factor $10^5$) for a model of instantaneous reionization. Red dashed line: scalar
        gravitational potential (set to unity deep in the radiation
        era). Blue continuous line: scalar part of baryon velocity. Blue dashed line:  scalar part of cold dark matter velocity. Green continuous line: scalar part of photons velocity. Green dashed line: scalar part of radiation anisotropic stress. All contributions are evaluated for the Fourier mode
        $k=0.01 \,{\rm Mpc}^{-1}$.} 
      \label{fig:eta'}
\end{figure}

The other terms in $C^\Y_{(i)(j)}$ involve quantities such as the temperature fluctuations $\Theta$, the
Compton parameter $y$ and the polarization tensor $\calP_{\mu\nu}$. Because metric and photon
perturbations do not grow, these quantities remain of the same order of the primordial potential $\Phi$
throughout the cosmic evolution. On the other hand, the baryon velocity $v_{(i)}$ after recombination
($\eta\gtrsim280\text{Mpc}$) grows like $(k \eta)\Phi$, as shown in Fig.~\ref{fig:eta'}. As a result, when
the baryons and the photons interact again during the era of reionization ($\eta\gtrsim5000\text{Mpc}$),
the $v^2$ term in \refeq{Cij} dominates over all the others in the
collision term, which can therefore be safely neglected. Furthermore,
we can ignore the contribution from $C^{{\Y}\, {\rm (L.O.)}}_{(i)(j)}$
coming from the time of recombination, because it was then suppressed
by the tight-coupling between photons and baryons by a factor $\propto
k/\tau'$~\cite{Peebles1970,Ma:1995ey,Pitrou:2010ai,CyrRacine:2010bk}. We remark that these assumptions
were numerically demonstrated in Ref.~\cite{Pitrouysky} in the similar context of the intensity spectral
distortion, where the contribution from recombination was shown to be suppressed by a factor ${\cal O}(10^2)$ compared to that from reionization.

Let us notice that because the distortion of polarization is, within our leading-order approximation, sourced only by quantities quadratic in first order perturbation theory, the tetrad basis can be considered for practical purposes to be calculated on the unperturbed background spacetime. From now on we will therefore omit the parentheses in $\Y_{(i)(j)}$ and the likes and consider the indices $i,j \ldots$ as trivial spatial comoving indices. 

In order to numerically compute the spectral distortion ${\Y}_{ij}$, we make use of the line of sight solution of the Boltzmann equation \cite{seljak:1996a}, which for \refeq{Y'} and in Fourier space formally reads\footnote{Our convention for the Fourier transformation is such that for every real-space quantity $A(\gr{x})$, we define its Fourier transform as $A(\gr{k}) = \int  {\rm d}^3 \gr{x}\,  e^{-i  \gr{k} \cdot \gr{x} }   A(\gr{x}).$}
\begin{equation}\label{IntegralForm}
  {\Y}_{ij}(\eta_0,k_i,n^{i}) \;=\; \int_{\eta_{\re}}^{\eta_0} \dd \eta \: \tau'\, e^{-\tau}\,
  e^{-\ii \,k_i\,
  n^{i} r(\eta)} \, C^{\Y}_{ij}(\eta,k_i,n^{i})
\end{equation}
where $\dd \tau(\eta) / \dd \eta \equiv -\tau'$, $\tau(\eta_0)=0$, $r(\eta) \equiv
\eta_0-\eta$ is the comoving distance from us, $\eta_0$ is the present conformal time and $\eta_{\re}$
denotes a time slightly before the beginning of reionization. At
leading order, replacing $C^{\Y}_{ij}$ by $C_{ij}^{{\Y} \,{\rm (L.O.)}}$ in
Eq.~\refeq{IntegralForm} provides an explicit integral
solution once the first-order velocity $v_{i}$ is known.

\subsection{Multipolar expansion of the polarization distortion}

We now proceed to a multipolar expansion of the distortion tensor
${\Y}_{ij}(\gr{k},\gr{\n})$ (we omit the mention of $\eta_0$ from now
on without ambiguity). We choose the $z$ axis along the direction of the Fourier vector $\gr{k}$ of interest. The traceless projected tensor ${\Y}_{ij}(\gr{k},\gr{\n})$ can then be expanded on the natural polarization basis $m_i^{\pm}$ as \cite{Hu1997,Hu2000}
\begin{equation}\label{defEB}
{\Y}_{ij}(\gr{k},\gr{\n}) \;=\; \sum_{\pm}\,\sum_{\ell=2}^{\infty} \sum_{m=-\ell}^\ell \left[\:E^\Y_{\ell m}(\gr{k}) \pm  i B^\Y_{\ell m}(\gr{k})\:\right] \: \frac{Y_{\ell m}^{\pm2}(\gr{\n})}{N_\ell}  \:m^{\pm}_{i}\,m^{\pm}_{j}
\end{equation}
where $N_{\ell} \equiv \ii^\ell \sqrt{(2\ell+1)/(4 \pi)}\,$,
$\,m_i^{\pm}  \equiv ({\hat e}^{\theta}_i \mp i {\hat e}^{\phi}_i
)/\sqrt{2}$ with standard spherical coordinates, and
where a hat on a vector indicates that it is of unit norm.

As for $v_i$ sourcing $\Y_{ij}$, it only contains a scalar mode in the standard cosmological model and reads 
\begin{equation}
v_i(\eta,\gr{k}) \;=\; -i\, {\hat k}_i \,\F(k,\eta)\, \Phi({\bf k})
\label{vi}
\end{equation}
where $\Phi({\bf k})$ is the primordial gravitational potential and $F(k,\eta)$ denotes the transfer function of the baryon velocity. From this, one can arrive at the multipolar decomposition of the leading order collision term (see Appendix \ref{vivj} for the derivation):
\begin{eqnarray}
C^{{\Y} \,{\rm (L.O.)}}_{ij}(\eta,\gr{k},\gr{\n})& \;=\; &\sum_{\ell,m,\pm}
E[C^{\Y}]_{\ell m}(\gr{k}) \: \frac{Y_{\ell m}^{\pm2}(\gr{\n})}{N_\ell} \:m^{\pm}_{i}\,m^{\pm}_{j} 
\label{multipole-C}
\end{eqnarray}
where
\begin{eqnarray}
E[C^{\Y}]_{\ell m}(\gr{k})& \;=\;& \delta_\ell^2\;{\cal K} \left\{\,S_{m}(\gr{{\hat k}}_1,\gr{{\hat k}}_2)
  \,\F(k_1,\eta)
  \,\F(k_2,\eta)\:\Phi(k_1)\Phi(k_2)\,\right\}
  \label{E-C}
\end{eqnarray}
and where
\begin{equation}
 \mathcal{K}\{\dots\} \;\equiv\; \int \frac{\dd^3\gr{k}_1
 \dd^3\gr{k}_2}{(2 \pi)^{3}}\; \delta_{\rm D}^{3}(\gr{k}_1+\gr{k}_2-\gr{k})\,\dots 
 \label{K}
\end{equation}
denotes the convolution operator.
The expression of the geometrical factor $S_{m}(\gr{{\hat k}}_1,\gr{{\hat k}}_2)$ can be found in Eq.~\refeq{Sm}. 

Similarly to the case of the standard polarization, the
(leading-order) collision term from Thomson scattering only contains
an electric part, \textit{i.e.} the magnetic part $B[C^{\Y}]_{\ell m}$
vanishes identically. However, free-streaming does produce an
observable magnetic pattern in the polarization distortion out of the
pure $E$-modes from the collision term. This can be seen by inserting
Eq.~\refeq{multipole-C} into the expression \refeq{IntegralForm} and by using the Rayleigh formula to expand $e^{-\ii\,k_i\,n^{i}r}$ into
spherical harmonics. The rules for the addition of
spherical harmonics can then be used to obtain expressions for the electric and magnetic
parts of the distortion tensor $\Y_{ij}$ analogous to the ones of the standard polarization \cite{Hu1997}:
\begin{eqnarray}
&&\frac{E^{\Y}_{\ell m}(\gr{k})}{2\ell+1}= {\cal K}\left\{\int_{\eta_\re}^{\eta_0}
  \dd \eta\, g(\eta) \,\epsilon_\ell^{(m)}[k r(\eta)] \,
S_{m}(\gr{{\hat k}}_1,\gr{{\hat k}}_2) \, F(k_1,\eta)F(k_2,\eta) \,  \Phi(k_1) \Phi(k_2)\right\}  \label{result-E} \msk
&&\frac{B^{\Y}_{\ell m}(\gr{k})}{2\ell+1}= {\cal K}\left\{\int_{\eta_\re}^{\eta_0}
  \dd \eta\, g(\eta) \,\beta_\ell^{(m)}[k r(\eta)] \,
S_{m}(\gr{{\hat k}}_1,\gr{{\hat k}}_2) \, F(k_1,\eta)F(k_2,\eta) \,  \Phi(k_1) \Phi(k_2)\right\}  \label{result-B}
\end{eqnarray}
where we defined the visibility function
\begin{equation}
g(\eta)=\tau' e^{-\tau}\,.
\label{g}
\end{equation}
The explicit expressions of the projection functions
$\epsilon_\ell^{(m)}$ and $\beta_\ell^{(m)}$ are reported for completeness in Appendix \ref{epsilon-beta-lm}.

\section{Angular power spectra of polarization spectral distortions} 
\label{Power-spectra}

In the previous section we have defined the multipolar decomposition Eq.~\refeq{defEB} of the Fourier components of the distortion tensor $\Y_{ij}$ and obtained the explicit expressions Eqs.~\refeq{result-E}-\refeq{result-B} of its electric and magnetic parts. In this section, we relate these quantities to correlation functions of real-space observables.

\subsection{Exact angular power spectra}

Out of the distortion tensor, one can extract the distortion Stokes parameters $Q^\Y$ and $U^\Y$ defined as
\begin{eqnarray}
\Y_{ij}({\bf x},{\bf {\hat n}}) &\;=\;& \sum_{\pm}\,({Q^\Y}\pm i\, {U^\Y})({\bf x},{\bf {\hat n}}) \:m_i^{\pm}\, m_j^{\pm} \,.
\end{eqnarray}
The latter can be decomposed onto spherical harmonics as
\begin{eqnarray}
({Q^\Y}\pm i {U^\Y})({\bf x},{\bf {\hat n}})&\;=\;& \sum_{\ell=2}^{\infty} \sum_{m=-l}^l (e^\Y_{\ell m}({\bf x})\pm i \,b^\Y_{\ell m}({\bf x})) \: Y_{\ell m}^{\pm2}(\gr{\n};{\bf {\hat e}})
\label{def-eb}
\end{eqnarray}
where in $\,Y_{\ell m}^{\pm2}(\gr{\n};{\bf {\hat e}})\,$ we have made it explicit that the spin-2 spherical harmonic is defined with respect to an (arbitrary) real-space axis ${\bf {\hat e} }$, contrary to the previous section in which this axis was aligned with the Fourier wavevector of interest. We then define the angular power spectra as
\begin{eqnarray}
C_\ell^{E^{\Y}} \equiv  \langle  |e^\Y_{\ell m}({\bf x}) |^2 \rangle \, \quad\text{and}\quad C_\ell^{B^{\Y}} \equiv  \langle  |b^\Y_{\ell m}({\bf x}) |^2 \rangle \;.
\label{def-Cl}
\end{eqnarray}
We do not consider the cross-correlation  $\langle e^\Y_{\ell m}({\bf x}) b^{\Y*}_{\ell'm'}({\bf x})   \rangle$ because it vanishes by parity conservation. In Appendix \ref{observables}, we give the relations between these power-spectra and angular correlation functions of spin-0 quantities built out of $Q^\Y$ and $U^\Y$.

Following standard calculations for the polarization, such as in Refs.~\cite{Hu1997,Durrer}, the power spectra can be expressed as
\begin{equation}
(2 \ell+1)^2 \,C_\ell^{E^\Y} \;=\; \frac{2}{\pi} \sum_{m=-2}^2 \int {\rm d} k \,k^2\,  {\cal Q}^{E^\Y}_{\ell m}(k) \;,
\label{Cl}
\end{equation}
and similarly for $C_\ell^{B^Y}$, where we defined
\begin{equation}
  \langle E_{\ell m}^{\Y}(\gr{k}) E_{\ell m'}^{\Y *}(\gr{k'}) \rangle \;=\;
  (2 \pi)^3 \,\delta_{\rm D}^3({\bf k}-{\bf k'}) \, {\cal Q}^{E^\Y}_{\ell m}(k)\, \delta_{m m'}\,.
\end{equation}

The crucial difference between our set-up and the one of the standard polarization is that in the latter case, $E_{\ell m}$ and $B_{\ell m}$ are linear in the Gaussian variable $\Phi$ whereas $E_{\ell m}^{\Y}$ and $B_{\ell m}^{\Y}$ in Eqs.~\refeq{result-E}-\refeq{result-B} are quadratic in the primordial potential $\Phi$, and hence involve a convolution operator. Using Wick theorem, one can find the following expressions:
\begin{equation}
  {\cal Q}^{E^\Y}_{\ell m}(k) \,=\,
  \frac{2 (2\ell +1)^2}{(2 \pi)^3 } \int  {\rm d} ^3 \gr{k}_1 P(k_1) P(k_2)
  \left| S_{m}(\gr{{\hat k}}_1,\gr{{\hat k}}_2)  \right|^2  \left|\int_{\eta_\re}^{\eta_0}
  \dd \eta \,g(\eta)\,\epsilon_\ell^{(m)}[k r(\eta)] F(k_1,\eta)F(k_2,\eta) \right|^2
 \label{E-Y}
\end{equation}
\begin{equation}
 {\cal Q}^{B^\Y}_{\ell m}(k) \,=\,
 \frac{2 (2\ell +1)^2}{(2 \pi)^3 } \int  {\rm d} ^3 \gr{k}_1 P(k_1) P(k_2)
 \left| S_{m}(\gr{{\hat k}}_1,\gr{{\hat k}}_2)  \right|^2  \left|\int_{\eta_\re}^{\eta_0}
 \dd \eta \,g(\eta)\,\beta_\ell^{(m)}[k r(\eta)] F(k_1,\eta)F(k_2,\eta) \right|^2
 \label{B-Y}
\end{equation}
where $P(k)$ is the primordial power spectrum of the potential $\Phi$, defined as 
\begin{equation}
\langle\Phi({\bf k})\,\Phi^*({\bf k'}) \rangle \;=\; (2 \pi)^3 \, \delta_{\rm D}^3({\bf k}-{\bf k'})\,P(k)\,,
\label{Primordial-Power}
\end{equation}
and where it is understood that $\gr{k_2}=\gr{k}-\gr{k_1}$. Because of statistical isotropy, the integrands in  Eqs.~\refeq{E-Y}-\refeq{B-Y} do not depend on the azimuthal angle of $\gr{k_1}$ around $\gr{k}$. Therefore the Fourier integrals there are only truly two-dimensional, that is, ${\rm d}^3 {\bf k}_1 = 2\pi \,k_1^2 \,{\rm d} k_1 \sin
\theta_{\gr{k}_1} \dd \theta_{\gr{k}_1}$, where $\theta_{\gr{k}_1}$ denotes the angle between $\gr{k_1}$ and $\gr{k}$. Let us note eventually that $ {\cal Q}^{E^\Y}_{\ell, -1}= {\cal Q}^{E^\Y}_{\ell ,1}$ and $ {\cal Q}^{E^\Y}_{\ell, -2}= {\cal Q}^{E^\Y}_{\ell ,2}$. For this reason, we simply call in the following the $m=1$ contribution to the spectra the sum of the $m=-1$ and $m=1$ contributions (and similarly for $m=2$).

\subsection{Limber approximation}\label{Limber}

In the small scale limit, one can simplify the exact results in
Eqs.~\refeq{Cl}-\refeq{E-Y}-\refeq{B-Y} by using the Limber
approximation \cite{Limber} (see also Refs.~\cite{LoVerde:2008re,Bernardeau2010}). Noting that the functions
$\epsilon_\ell^{(m)}$ and $\beta_\ell^{(m)}$ are built out of the
spherical Bessel functions $j_\ell$ and their derivatives (see Appendix
\ref{epsilon-beta-lm}), and using that for a slowly varying function with respect to the oscillations of the $j_l$'s,
\begin{equation}
\int_0^{\infty} f(x)\,j_{\ell}(x) \;=\; \sqrt{\frac{\pi}{2 \nu}} \, f\left(\nu\right)+{\cal O}\left(\frac{1}{\nu^2}\right)\,,
\end{equation}
where $\nu=\ell+\frac12$\,, one obtains the leading-order results
\begin{eqnarray}\label{EqLimberEE}
  C_{\ell\, {\rm Limber}}^{E^{\Y}} &\;=\;&
  \frac{1}{4\,(2 \pi)^2} \,\int_0^{r_\re} \frac{{\rm d} r }{r^2}\, k_1^2 \,{\rm d} k_1
  \sin\theta_{\gr{k}_1} \dd \theta_{\gr{k}_1} \,P(k_1) P(k_2)
  \,\left[\,g(\eta)\,F(k_1,\eta)\,F(k_2,\eta) \,\right]^2 \nonumber \msk
  && \qquad \qquad  \times \left(\,3\,|S_{0}(\gr{{\hat k}}_1,\gr{{\hat k}}_2)|^2 \,+\,
  |S_{2}(\gr{{\hat k}}_1,\gr{{\hat k}}_2)|^2 \,\right) 
\end{eqnarray}
and
\begin{equation}
  C_{\ell\, {\rm Limber}}^{B^{\Y}} \;=\;
  \frac{1}{(2 \pi)^2} \,\int_0^{r_\re} \frac{{\rm d} r }{r^2}\, k_1^2 \,{\rm d} k_1 \sin
  \theta_{\gr{k}_1} \dd \theta_{\gr{k}_1} \,P(k_1) P(k_2) \,\left[\,g(\eta )\,F(k_1,\eta)\,F(k_2,\eta) \,\right]^2 \:|S_{1}(\gr{{\hat k}}_1,\gr{{\hat k}}_2)|^2 
\end{equation}
where $\gr{k_2}=\gr{k}-\gr{k_1}$, $k r =\ell+\frac12$ and
$\eta=\eta_0-r$. Note that within this approximation, modes with
$m=\pm1$ (respectively with $m=0$ and $m=\pm2$) do not contribute to
$C_{\ell\ }^{E^{\Y}}$ (respectively to $C_{\ell\ }^{B^{\Y}}$). We will
see in the next subsection that the above formulae, which are
numerically much easier to evaluate than the exact results, do indeed
provide an excellent approximation for $\ell \gtrsim 10$.

\begin{figure}[!htb]
  \includegraphics[width=0.5\textwidth]{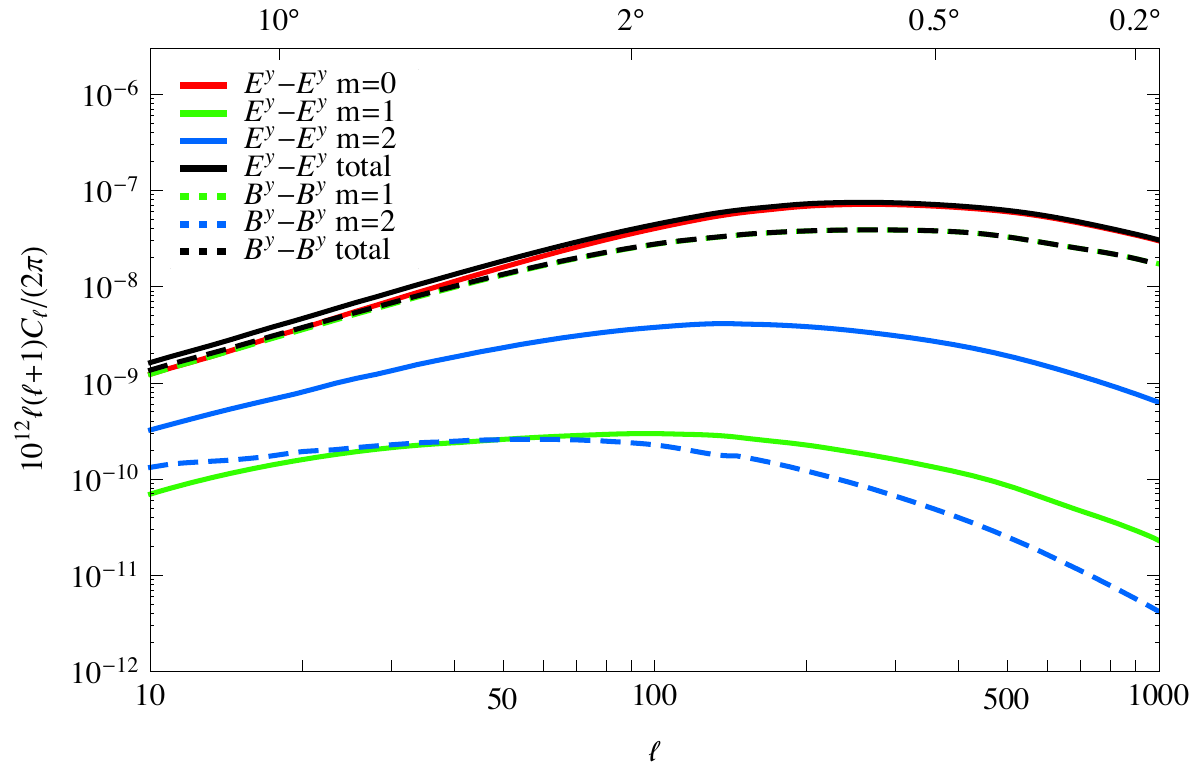}\includegraphics[width=0.5\textwidth]{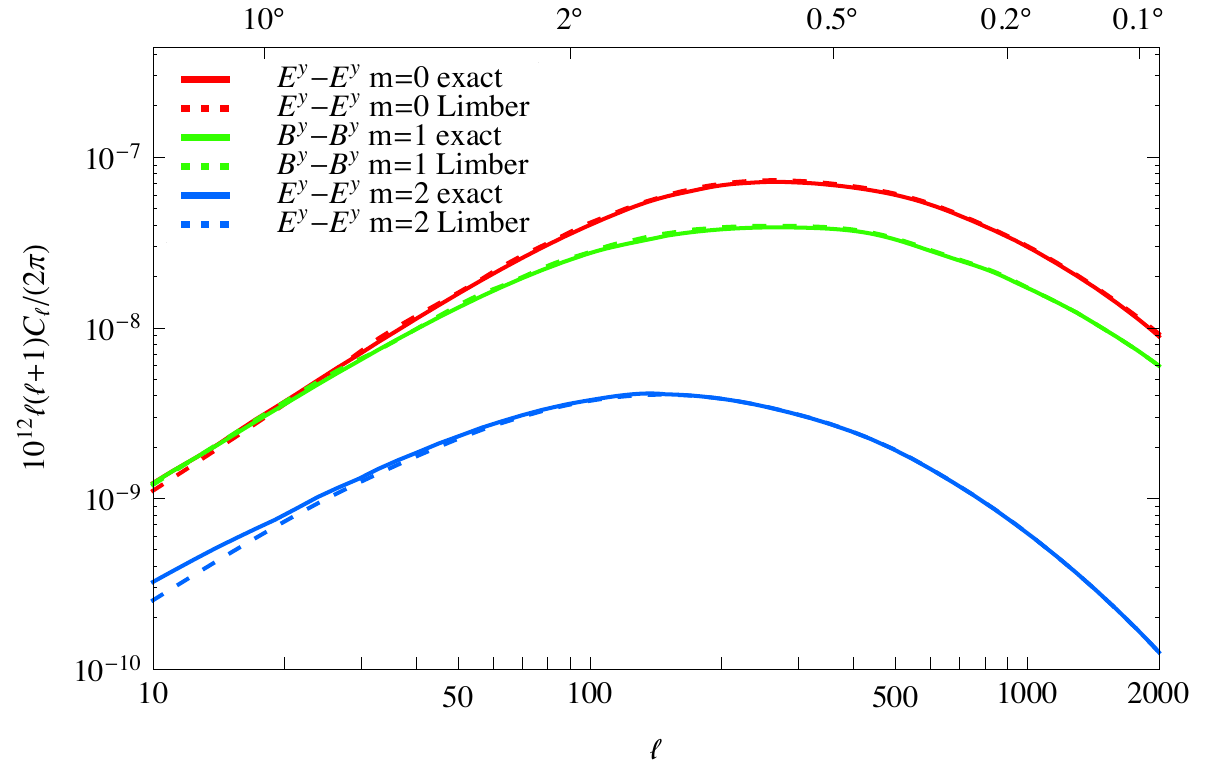}
 \caption{{\it Left.} Continuous black line: total $E^y$-modes spectrum;
   dashed black line: total $B^y$-modes spectrum. Individual
   contribution to the $E^y$-modes for $m=0,1,2$ are respectively in
   red, green and blue continuous lines. Individual contributions to
   $B^y$-modes for $m=1,2$ are respectively in green and blue dashed lines. 
 {\it Right.}  We compare the exact spectra to their Limber
approximations. Red: $m=0$ contribution; exact formula in
continuous line and Limber approximation in dashed line for $E^y$-modes
spectrum. Green: $m=1$ contribution; exact formula in
continuous line and Limber approximation in dashed line for $B^y$-modes
spectrum. Blue: $m=2$ contribution; exact formula in
continuous line and Limber approximation in dashed line for $E^y$-modes spectrum.
}      
\label{fig:ClEEBBdisto}
\end{figure}

\begin{figure}[!htb]
  \includegraphics[width=0.5\textwidth]{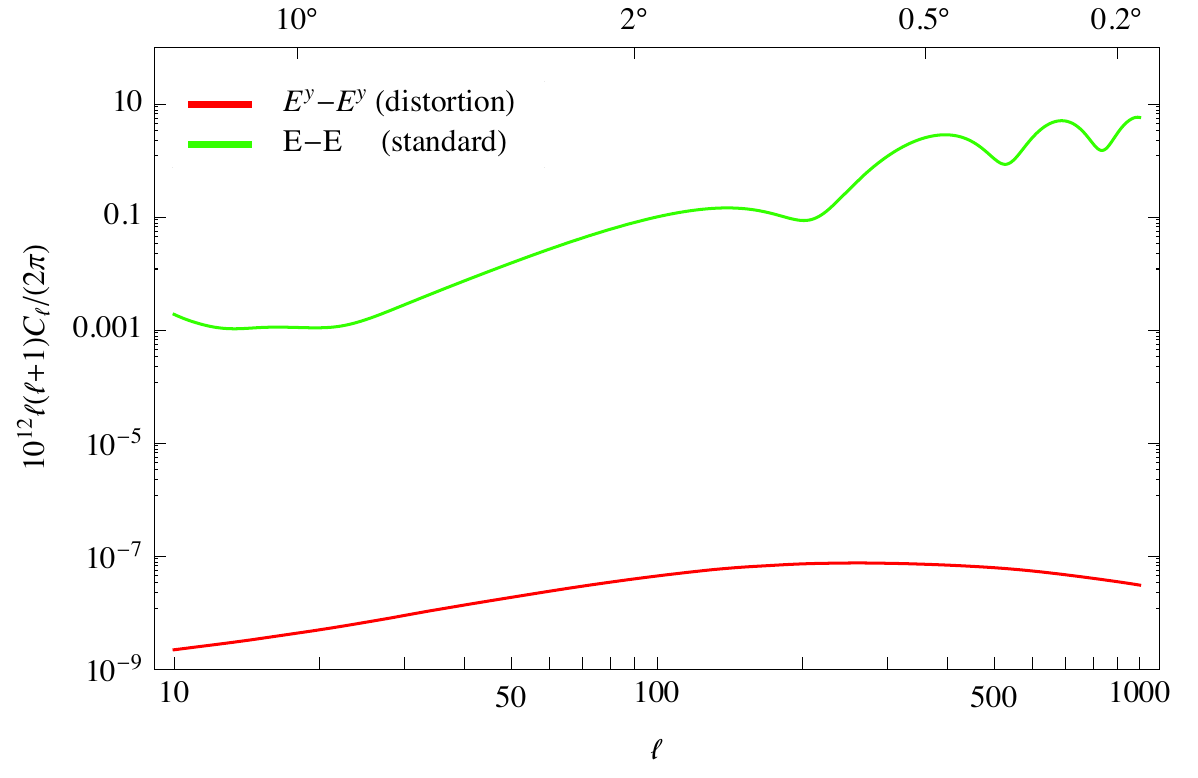}
\includegraphics[width=0.5\textwidth]{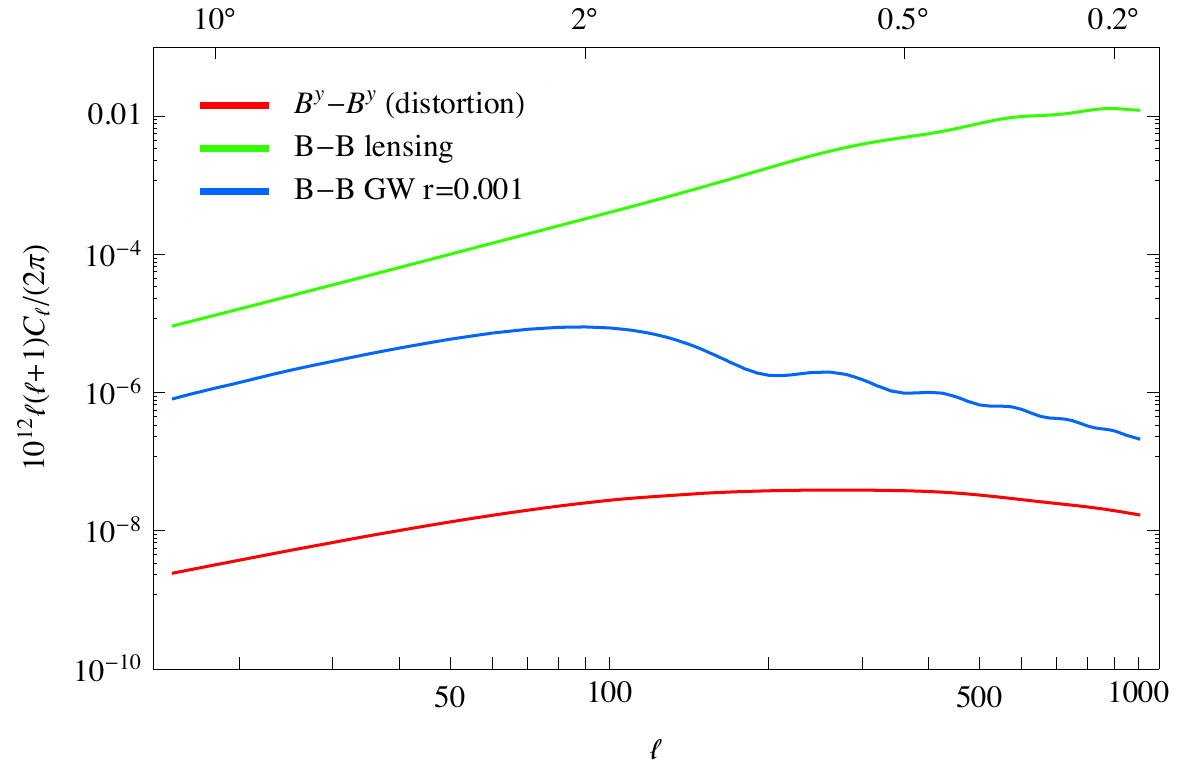}
 \caption{{\it Left.} Red: the $E^y$-modes spectrum of the
   $y$-type distortion (this article) compared to the standard $E$-mode polarization spectrum (green). {\it Right.} The $B^y$-modes spectrum of the
   $y$-type distortion (this article) in red compared to the standard $B$-modes spectrum induced by weak lensing (in green) and induced by primordial gravitational waves with tensor to scalar ratio $r=0.001$ (in blue).}

      \label{fig:ClEEBB}
\end{figure}

\section{Numerical results and discussion}\label{SectionNumResults}
\label{results}

We compute the power spectra of the polarization distortion $E^y$- and
$B^y$-modes by numerically evaluating the integrals in
Eqs.~\refeq{Cl}-\refeq{E-Y}-\refeq{B-Y} with the second-order
Boltzmann code \textsf{SONG}~\cite{Pettinari:2013he,BmodesCFGP} (Second Order
Non-Gaussianity), using Planck cosmological
parameters~\cite{Ade:2013zuv}. Unless otherwise stated, we
  consider a simple model of instantaneous reionization (see
  Fig.~\ref{fig:eta'}). We quantify the effects of an extended period of reionization in the last paragraph.

\paragraph{Signal of polarization distortion from reionization} In the left panel of Fig.~\ref{fig:ClEEBBdisto}, we plot the total power spectra for the $E^y$- and $B^y$-modes together with their individual contributions $m=0,1,2$. On the right panel, we compare the Limber approximations to the exact formulas, and see that the agreement is excellent, with the error going as $1/\ell$.
The $B^y$-modes are of the same order of magnitude as the
$E^y$-modes, since the sources are the same apart from geometrical
factors. The signal is very smooth with no baryon acoustic oscillation structure since
the baryon velocity is almost completely
similar to the dark matter velocity during reionization (see Fig.~\ref{fig:eta'}). It peaks around $\ell=280$, with $\ell(\ell+1)C_\ell/(2 \pi) \simeq 7.\,10^{-20}$, its amplitude being controlled by the optical depth to reionization. When compared with
the cluster contribution reported in Ref.~\cite{Baumann:2002es}, it is one order of
magnitude larger at this scale. This can be understood from
simple arguments. First, the local velocities of baryons in the
intergalactic medium or of the point sources such as galaxy clusters are expected to be of the
same order.
Second, it is true that the baryon density is higher inside clusters due to the
non-linear collapse of matter. However, when averaged at a given
redshift over a volume of typical size larger than
the typical intercluster scale, the number of electrons should be of the same order as the one found from the linear
description. Provided we consider angular scales subtending transverse
distances that are larger than the typical intercluster scale, we thus expect the contribution from the intergalactic
medium computed in our mildy non-linear formalism (that is second order)
to account correctly for the contribution due to the kinetic
motion of clusters. With a typical intercluster distance of $20\,\,{\rm Mpc}$, seen at a redshift $z=1$, that is at a comoving distance
of approximately $3400\,\,{\rm Mpc}$, it corresponds to angular scales
larger than $20'$, that is to multipoles $\ell$ smaller than roughly $500$. Finally, one should
however take into account the fact that the contribution from the intergalactic medium develops from the
beginning of reionization (around $z=12$) onwards --- when its contribution is
the largest --- whereas the contribution coming from the galaxy clusters peculiar velocity starts to contribute at
much later times, when galaxy clusters have formed. Hence, the total
contribution integrated over all redshifts of the intergalatic medium
computed with second-order perturbation theory should be larger than the clusters contribution, at least in the range
where the mildly non-linear description accounts correctly for
the clusters contribution, that is for $\ell \lesssim 500$. 

Finally, note that our results are in agreement with the ones of Hu in Ref.~\cite{Hu:1999vq}, who performed a similar calculation within the Limber approximation. Let us remark however that, as we have explained in \S~\ref{LOsolution}, this reference considers the brightness of the polarized signal by integrating over its spectral dependence. On the contrary, the polarization tensor is separated here between a standard component and a proper spectral distortion component [see Eq.~\refeq{defyPbase}], that future observations have the potential to disentangle.

\begin{figure}[!htb]
  \includegraphics[width=0.5\textwidth]{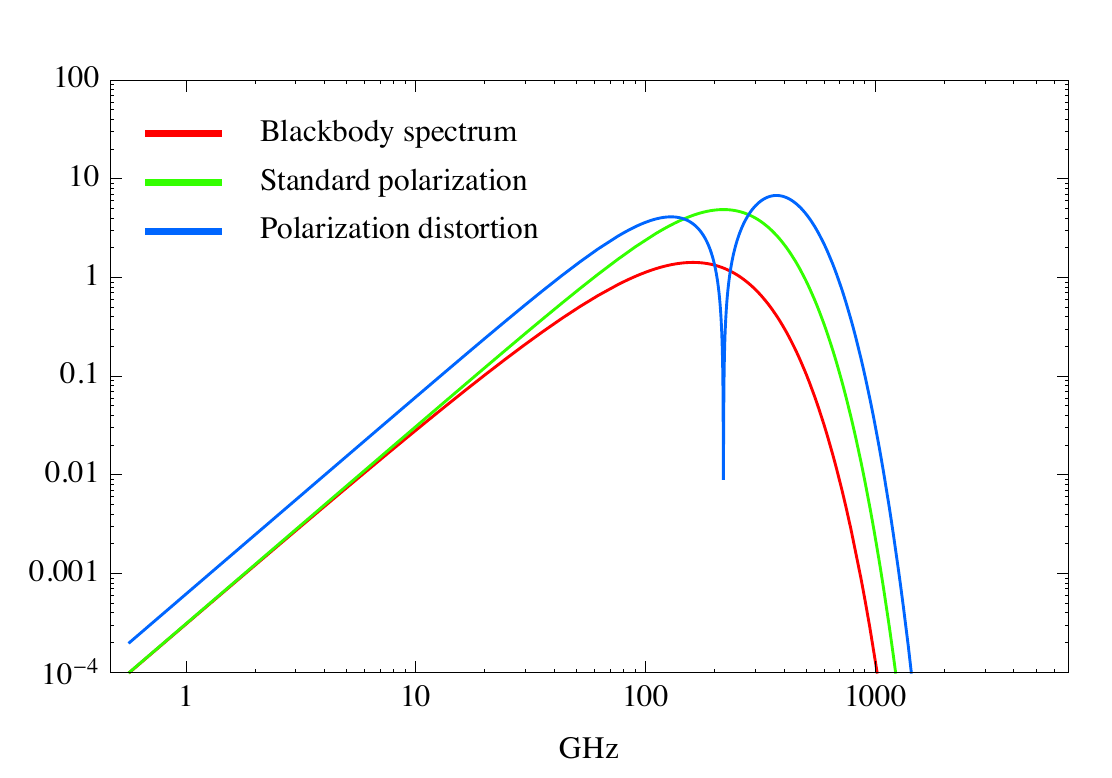}
\includegraphics[width=0.5\textwidth]{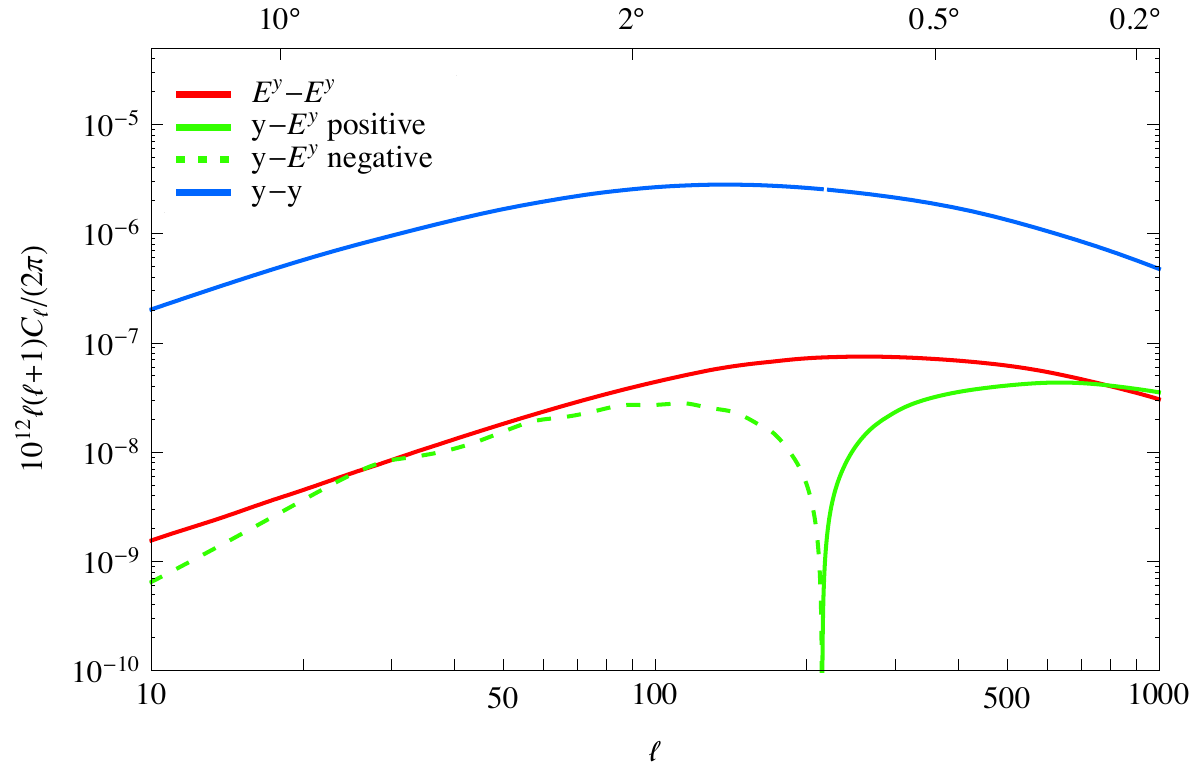}
      \caption{{\it Left. }Red: intensity brightness for  a blackbody spectrum
        [$E^3 I_{\rm BB}(E/T_0)$]. Green: brightness of the standard polarization [$E^3 \partial I_{\rm
          BB}(E/T_0) /\partial \ln E$]. Blue: brightness of the
        $y$-type distortion  [$E^3 {\cal D}^2_E I_{\rm
          BB}(E/T_0)$] (see Eqs.~\refeq{defy} and \refeq{defyPbase}). All curves have been divided by $T_0^3$ to make them dimensionless, with $T_0=2.73\,{\rm K}$. The blackbody brightness peaks at
        $160\,{\rm GHz}$, while the y-type distortion brightness peaks
        at $370\,{\rm GHz}$.       
{\it Right.} Red: the angular power spectrum of polarization
        distortion $E^y$-modes; Green: the cross
        correlation of polarization distortion $E^y$-modes
        with $y$-type intensity distortion (continuous when positive
        and dashed when negative); Blue: the angular power spectrum of
        intensity $y$-type distortion.
}
    \label{fig:SpectralFunctions}
\end{figure}

\paragraph{Comparison with the standard polarization} In Fig.~\ref{fig:ClEEBB} we also compare the power spectra of the $E^y$- and $B^y$-modes
of the $y$-type distortion with the spectra of the standard $E$- and $B$-modes of
polarization. Since the $y$-type polarization distortion is a
second-order effect, its power spectrum is expected to be much smaller than the
standard polarization, as it involves two powers of the
primordial power spectrum. The latter being of
order $10^{-10}$, one would expect the spectrum of the $E^y$-modes to be roughly $10$ order of magnitude smaller than the one of the standard
$E$-modes. One could even think it would be $12$ orders of
magnitude smaller if we take into account the fact that only $10\,\%$
of the visibility function $g(\eta)$ contribute in the reionization era.
However, one should take into account  the facts that {\it i)} the standard $E$-modes are already suppressed by tight-coupling during recombination and {\it ii)} the sources \refeq{Cij} are
quadratic in the linear baryon velocity, which in turn is boosted by a factor $k\eta$
with respect to the primordial fluctuations (see Fig.~\ref{fig:eta'}).
We find indeed that the spectrum of the polarization distortion $E^y$-modes is
approximately only $6$ orders of magnitude smaller than the one of the standard
$E$-modes at $\ell=200$. Furthermore, the spectrum of the polarization distortion $B^y$-modes is
two orders of magnitude smaller than the contribution from primordial
gravitational waves with tensor to scalar ratio $r=0.001$. However, one should remember that the
spectral shape in the $y$-type polarization distortion signal is very different from the
standard polarization, and that there is more signal in the former than in
the latter.
This can be seen from the left panel of Fig.~\ref{fig:SpectralFunctions}, where we show the
different spectral shapes of polarization, namely the standard signal and the
$y$-type distortion, together with a blackbody spectrum for
comparison. The brightness for the $y$-type distortion peaks at higher
frequency than the brightness of the standard signal, and for nearly all frequencies it is larger in absolute value.

\begin{figure}[!htb]
  \includegraphics[width=0.5\textwidth]{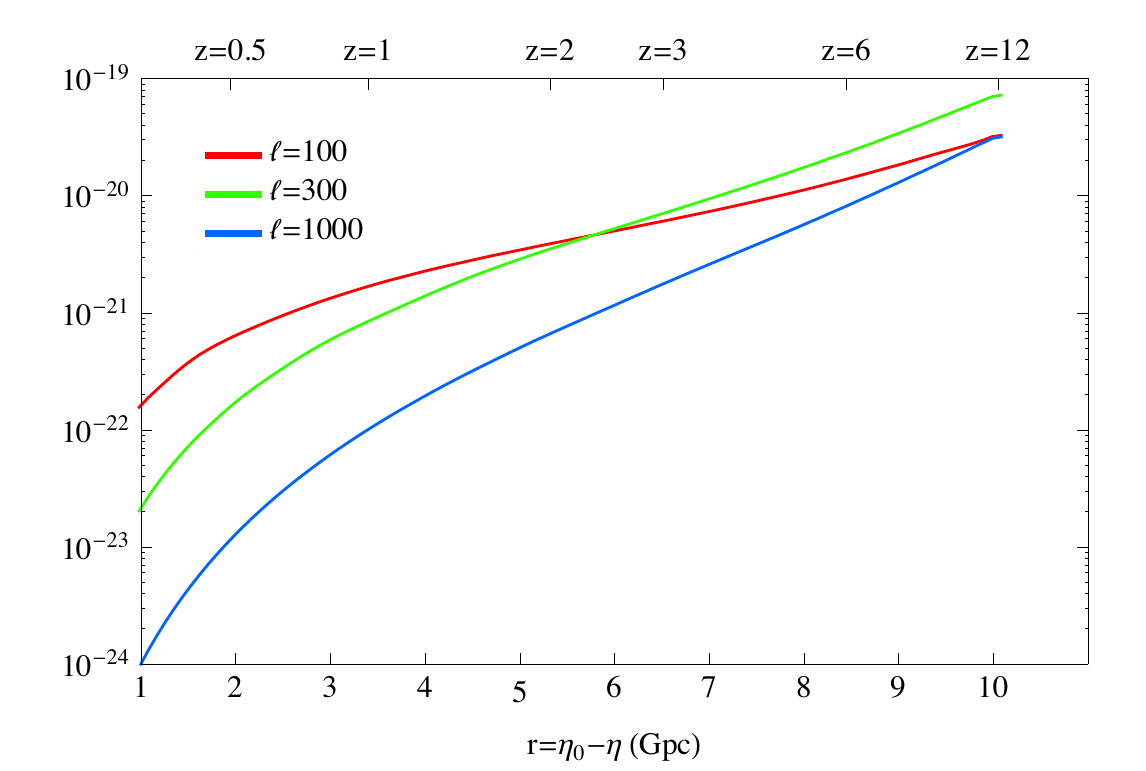}
\includegraphics[width=0.5\textwidth]{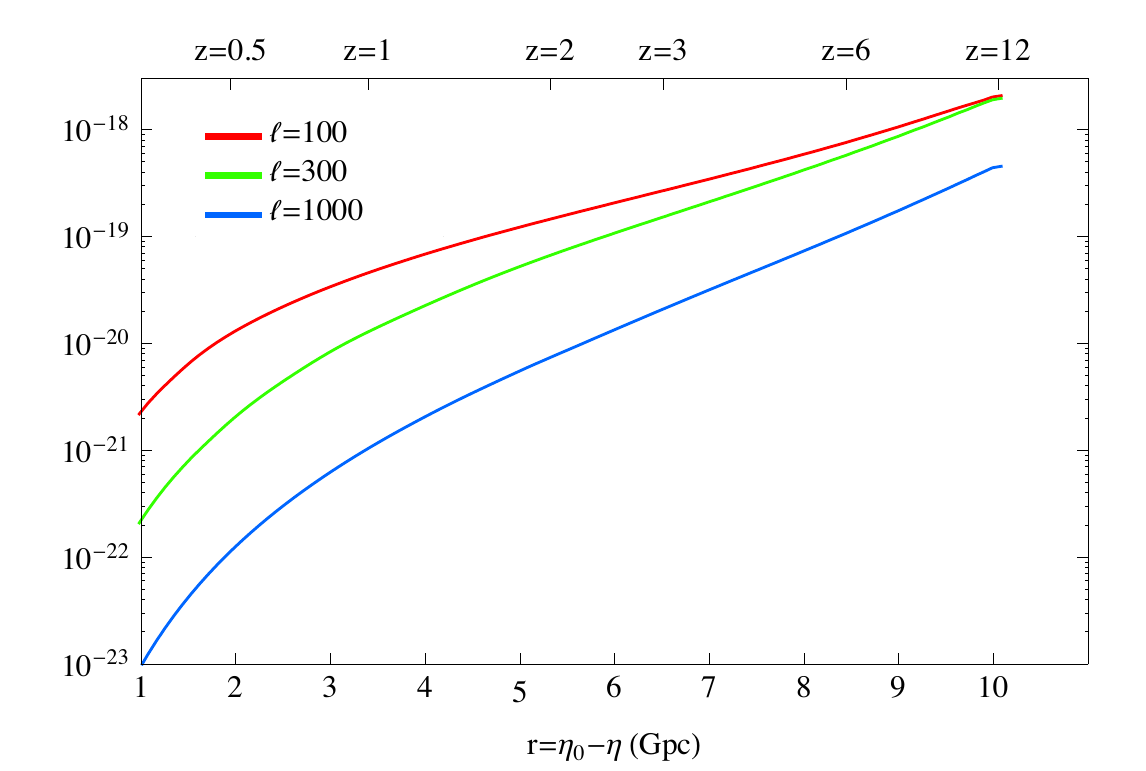}
      \caption{{\it Left.} From the Limber
        approximation~(\ref{EqLimberEE}) for the polarization distortion
        $E^y$-modes spectrum, we plot the integrand of the outer
        integral on the comoving distance $r$ [multiplied by $\ell (\ell+1)/(2 \pi)$], for $\ell=100$
        (red), $\ell=300$ (green) and $\ell=1000$ (blue).
{\it Right.} From the Limber approximation~(\ref{EqLimberyy}) for
the intensity $y$-type distortion spectrum, we plot the integrand of the outer integral on the comoving distance $r$ [multiplied by $\ell (\ell+1)/(2 \pi)$], for $\ell=100$
        (red), $\ell=300$ (green) and $\ell=1000$ (blue).
}
    \label{fig:TimeContrib}
\end{figure}

\paragraph{Improving the detectability with cross-correlations} To improve the hopes of detection of the $y$-type polarization distortion, one can consider its cross-correlation with the $y$-type intensity distortion, the latter being
larger in magnitude~\cite{Pitrouysky} and with more signal-to-noise. One might be concerned that the intensity $y$-type
distortion is not dominated by the contribution from the non-linear
kSZ effect of the intergalatic medium, but rather by the tSZ effect from unresolved
clusters. However the latter will contribute only at low redshifts
when galaxy clusters have formed, whereas the distortion from the non-linear kSZ
effect due to the intergalactic medium, both in intensity and polarization, contributes mostly at the beginning of
reionization around $z\simeq 12$, where the visibility function is the
largest. This can be seen more rigorously by examination of the integrand of the Limber
approximation [e.g. Eq.~(\ref{EqLimberEE})] for the angular spectrum of
polarization distortion $E^y$-modes. Once the Fourier modes integral has been performed,
there remains only one integral on the comoving distance $r$, and by
plotting the corresponding integrand in Fig.~\ref{fig:TimeContrib} for several multipoles, we
find indeed that it is larger for high redshifts. The same result is
found for the intensity $y$-type distortion spectrum.
Apart from the very large angular scales, there would thus be no correlation between the tSZ
signal and the intergalactic contribution to the non-linear kSZ
effect.
In Appendix~\ref{Cross}, we summarize the formalism necessary
to compute the correlation between the $y$-type polarization distortion and the $y$-type intensity
distortion.
We have evaluated it numerically using \textsf{SONG} and plotted the results thus obtained
in the right panel of Fig.~\ref{fig:SpectralFunctions}.
Unfortunately, it appears that the cross-correlation of the $E^y$-modes polarization distortion with the intensity distortion is
actually not larger, but on the contrary slightly smaller in magnitude, than the auto correlation of the $E^y$-modes. However, to improve the detectability of our signal, correlating it with the $y$-type intensity distortion is not the only possibility. As we have mentioned in \S.~\ref{LOsolution}, the non-linear kSZ effect not only generates a spectral distortion component but also a similar contribution to the standard component ${\cal P}_{\mu \nu}$ of the polarization tensor. More precisely, the latter obeys an evolution equation similar to Eq.~\refeq{Y'} in the tetrad basis, with a collision term containing 4 times the contribution \refeq{Cij}. This second-order effect generates a correction to the standard $E$- and $B$-modes which is thus simply four times their spectrally distorted counterparts \refeq{result-E} and \refeq{result-B}, leading to correlations $\langle E^{\rm st} E^{y *} \rangle= 4 \langle E^{y} E^{y *} \rangle$ and $\langle B^{\rm st} B^{y *} \rangle= 4 \langle B^{y} B^{y *} \rangle$  where {\rm st} stands for the standard $E$- and $B$-modes (the leading order contributions to the latter do not correlate with the polarization distortion in a Gaussian universe, as the corresponding correlation is proportional to the bispectrum of the primordial gravitational potential).

\begin{figure}[!htb]
  \includegraphics[width=0.5\textwidth]{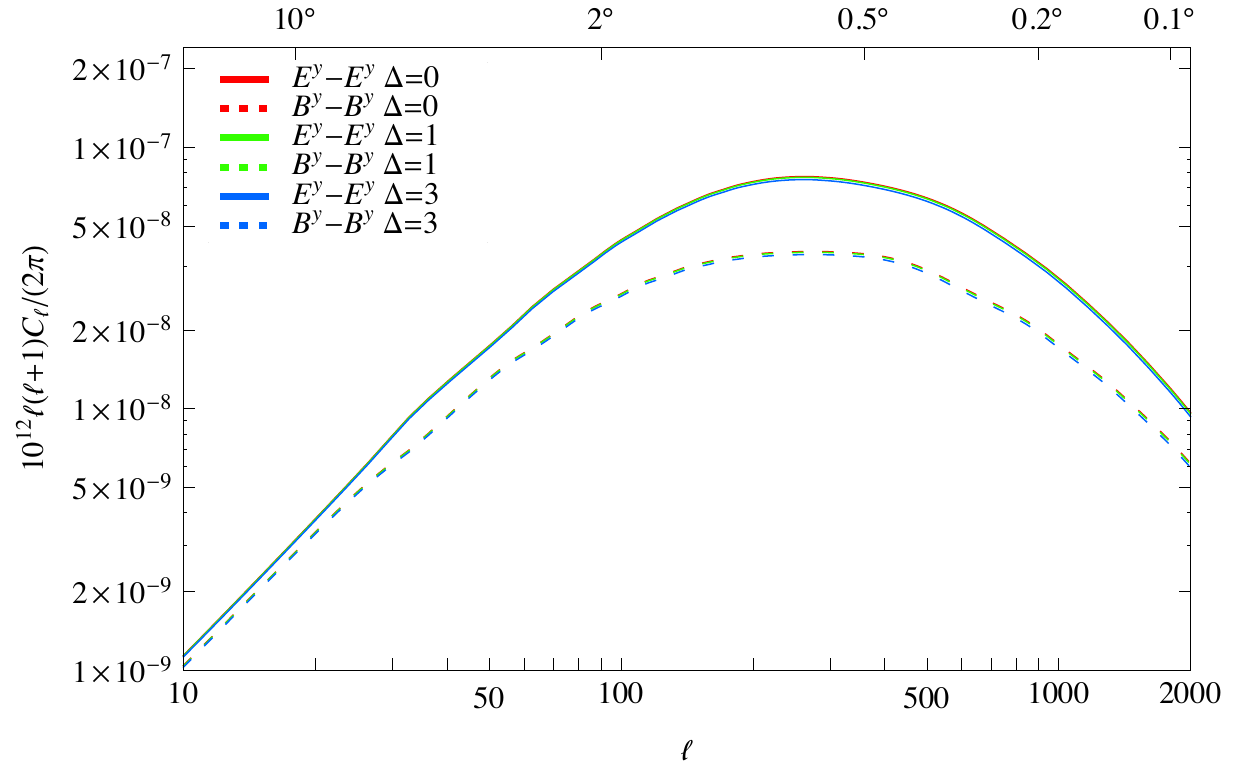}
\includegraphics[width=0.5\textwidth]{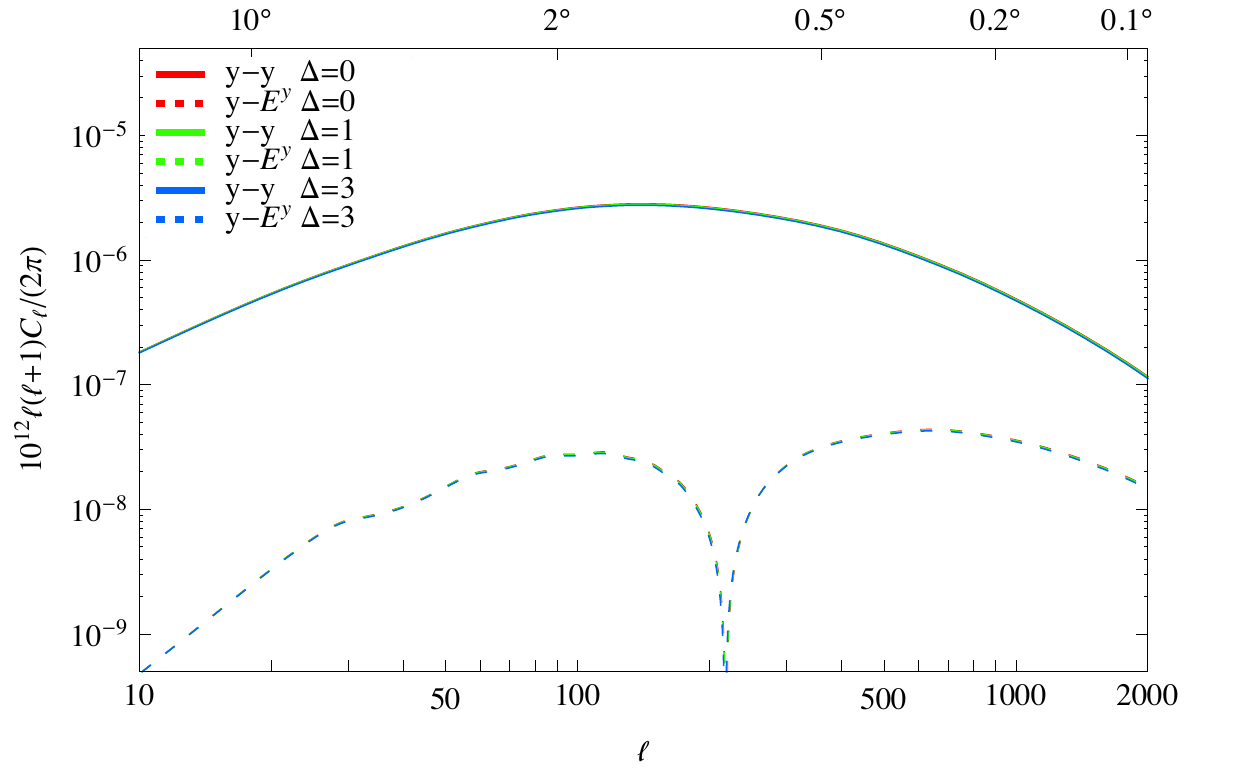}
 \caption{ Effects of an extended period of reionization. {\it Left.}
   The total $E^y$-modes spectrum (continuous lines) and the total
   $B^y$-modes spectrum (dashed lines) for the model of instantaneous
   reionization (in red) and for the model of extended reionization
   \refeq{xe} with the same value of $z_r$ but with $\Delta=1$ (green) and $\Delta=3$ (blue). {\it Right.} The angular power spectrum of intensity $y$-type distortion (continuous lines) and the cross
        correlation of polarization distortion $E^y$-modes
        with $y$-type intensity distortion (dashed lines) for the
        model of instantaneous reionization (in red) and for the model
        of extended reionization \refeq{xe} with the same value of
        $z_r$ but with $\Delta=1$ (green) and $\Delta=3$ (blue).
}      
\label{fig:ClEEBBVariableDelta}
\end{figure}

\paragraph{Effects of an extended period of reionization} The previous results were derived using a simple model of instantaneous reionization. In order to quantify to which extent a more realistic extended period of reionization modifies our signal, we consider the simple two-parameter model for the ionization history that is currently the default parameterization of the CAMB code \cite{Lewis:1999bs,Lewis:2008wr} and one of the two possible parameterizations of the CLASS code \cite{class,Blas:2011rf}. In this model, the number of free electrons per hydrogen atom $x_e$ is given by
\begin{equation}
x_e(z)\equiv \frac{n_e(z)}{n_H(z)}= \frac{f}{2}\, \Bigg\{1+{\rm tanh}\left[\frac{(1+z_r)^{3/2}-(1+z)^{3/2}}{\Delta} \right]  \Bigg\}\,,
\label{xe}
\end{equation}
where $f=1+n_{He}/n_H \sim 1.08$, $z_r$ is the redshift at which the hydrogen is half neutral and `the duration of reionization' $\Delta$ is the width of a tanh function that describes the time evolution of $x_e$. Note that the motivation for this simple model is merely mathematical: it is built such that the total optical depth is independent of $\Delta$ and thus coincides with the one of a model of instantaneous reionization, corresponding to the limit $\Delta \to 0$. In Fig.~\ref{fig:ClEEBBVariableDelta} we plot the modifications
induced by an extended period of reionization corresponding to $\Delta=1$ and $\Delta=3$, together with the results for the reference instantaneous model, for the total $E^y$- and $B^y$-modes spectra (left), and for the $y$-type intensity distortion spectrum and its cross-correlation with the $E^y$-modes (right). From these plots, it is readily apparent that all these spectra are nearly insensitive to the width of the reionization transition, the difference between the spectra corresponding to $\Delta=0$ and $\Delta=3$ being of 2\% at the location of the peak (and nearly insensitive to the multipole $\ell$). This shows that the physical processes studied in this paper constitute a probe of the total optical depth to reionization only.

\section{Conclusions} 
\label{sec:conclusion}

CMB anisotropies measurements have now entered a high precision era in
angular resolution. In addition, future CMB experiments aim to measure with great accuracy the deviation from blackbody nature of these anisotropies. This will open a new field of research and a new observational window into out-of-equilibrium and energy injection phenomena that cannot be probed otherwise, as well as it will add possibilities to set constraints on cosmological parameters. In particular, the angular correlations of temperature fluctuations, in
two point function or higher order statistics, are not enough if we
want to capture optimally the spectral information. Since CMB
experiments measure both the intensity and the linear polarization,
this spectral characterization has to be performed for both
signals. Indeed, the main physical effect responsible for the spectral
distortions of polarization is different from the one responsible for
the intensity distortion. In the former case, the non-linear kSZ effect is the dominant contribution, whereas the thermal SZ effect
dominates largely the intensity distortion. The cosmological
information gathered from spectral distortions in polarization thus
provides independent information. 

In this article, we have computed the angular power spectrum of spectral distortions in polarization generated by the
intergalactic medium. It is sourced by the difference between the
velocities of electrons and photons, and its dominant
contribution is therefore coming from the epoch of reionization. Even
though the signal is very small, it is guaranteed in the
standard cosmological model, and since its spectral shape is very
different from the one of the standard
polarization of the CMB, it is in principle possible to disentangle them. Spectral distortions in polarization are mainly sensitive to the optical depth to reionization
$\tau_{\rm re}$ for which it could provide an independent estimation.
Finally, we have found that most of the signal is on intermediary
scales with an angular power spectrum which does not have baryon
acoustic oscillation features. The numerical computation were performed with
the second-order code \textsf{SONG} and we have shown that a Limber approximation is sufficient to
compute the theoretical expectation of a given model for these scales
of interest.

\subsection*{Acknowledgements} 

We would like to thank J.-P. Uzan and B. Wandelt for discussions related to the topic of this paper, as well as the anonymous referee whose useful comments helped us to improve it. This work was supported by French state funds managed by the ANR
within the Investissements d'Avenir programme under reference
ANR-11-IDEX-0004-02. G. Pettinari (respectively C. Fidler) acknowledges support from the UK Science and Technology Facilities Council grant number ST/I000976/1 (respectively grant numbers ST/K00090/1 and ST/L005573/1).

\appendix

\section{Multipolar expansion of the collision term}
\label{vivj}

In this appendix, we show how to obtain the multipolar expansion of
\begin{equation}
C^{{\Y}\, {\rm (L.O.)}}_{ij}(\gr{k},\gr{\n})=-\frac{1}{10} {\cal K} \left\{  [v_{i}(\gr{k}_1) v_{j}(\gr{k}_2)]^{\rm TT} \right\}
\end{equation}
where we omit to mention the obvious time-dependence, and we recall that $ {\cal K}$ denotes the convolution operator \refeq{K}. Writing
\begin{eqnarray}
C^{{\Y} \,{\rm (L.O.)}}_{ij}(\gr{k},\gr{\n})& =&\sum_{\pm,\ell,m} \frac{1}{N_{\ell}} [E[C^{\Y}]_{\ell m}(\gr{k}) \pm  i B[C^{\Y}]_{\ell m}(\gr{k})] Y_{\ell m}^{\pm2}(\gr{\n}) m^{\pm}_{i}m^{\pm}_{j} \,,
\end{eqnarray}
we can extract, using $m_i^{\lambda} m^{\lambda' i}=1-\delta^{\lambda,\lambda'}$ and $n^i m_i^{\pm}=0$:
\begin{equation}
\sum_{\ell m}\, [E[C^{\Y}]_{\ell m}(\gr{k}) \pm  i B[C^{\Y}]_{\ell m}(\gr{k})] \frac{Y_{\ell m}^{\pm2}(\gr{\n})}{N_{\ell}}=-\frac{1}{10} {\cal K} \left\{  m^{\mp i} m^{\mp j}  v_{i}(\gr{k}_1) v_{j}(\gr{k}_2)   \right\}\,.
\end{equation}
Now, we use Eq.~\refeq{vi} to write
\begin{equation}
v_{i}(\gr{k}_1) n^{i}=\frac{1}{N_1} \sum_{m=-1}^{1} Y_{1m}(\gr{\n}) v_m(\gr{k}_1)
\label{vini}
\end{equation}
where
\begin{equation}
v_m(\gr{k_1})=\sqrt{\frac{4 \pi}{3}} Y_{1m}^*(\gr{{\hat k_1}}) F(k_1,\eta) \Phi(k_1)
\end{equation}
and where we recall that the spherical harmonics are defined with the $z$ axis aligned along the direction of the Fourier vector $\gr{k}$ of interest. Equipped with the projection Eq.~\refeq{vini} of $v_i$ along the local normal vector $n^{i}$, we obtain its projection on the local polarization basis by differentiating this relation:
\begin{equation}
m^{\pm j} \nabla_{j}( v_{i}(\gr{k}_1) n^{i}  )= v_{i}(\gr{k}_1) m^{\pm i}\,,
\end{equation}
where $\nabla_i$ is the covariant derivative on the unit-sphere. With
\begin{equation}
m^{\pm j} \nabla_{j}( Y_{1m} )=\pm  Y^{\mp 1}_{1m}\,,
\end{equation}
we therefore obtain
\begin{equation}
m^{\mp i} m^{\mp j}  v_{i}(\gr{k}_1) v_{j}(\gr{k}_2)= \frac{1}{N_1^2} \sum_{m\, n} v_{m}(\gr{k}_1) v_{n}(\gr{k}_2) Y^{\mp 1}_{1m} Y^{\mp 1}_{1n}
\end{equation}
Noting eventually that
\begin{equation}
Y^{\pm 1}_{1m} Y^{\pm 1}_{1n}=\frac{3}{2 \sqrt{5 \pi}}\left<1,m;1,n | 2,m+n \right>  Y^{\pm 2}_{2,m+n}\,,
\end{equation}
where $\left<l_1,m_1;l_2,m_2 | l,m_1+m_2 \right>$ denotes the corresponding Clebsch-Gordan coefficient, we finally identify that $B[C^{\Y}]_{\ell m}=0$ and
\begin{eqnarray}
E[C^{\Y}]_{\ell m}(\gr{k})& =& \delta_\ell^2 {\cal K} \left\{S_{m}(\gr{{\hat k}}_1,\gr{{\hat k}}_2)
  \F(k_1,\eta)
  \F(k_2,\eta)\Phi(k_1)\Phi(k_2)\right\}
\end{eqnarray}
where
\begin{equation}
S_{m}(\gr{{\hat k}}_1,\gr{{\hat k}}_2)=-\frac{\pi}{15} \sqrt{ \frac{2}{3} }\sum_{n=-1}^1
\alpha_{n,m}  \left( Y_1^{m-n}(\gr{\hat{k}_1}) Y_1^{n }(\gr{\hat{k}_2}) \right)^*
\label{Sm}
\end{equation}
with the notation
\begin{equation}
\alpha_{0,m}\equiv
\sqrt{(4-m^2)}, \qquad \alpha_{\pm1,m}   \equiv  \sqrt{(2 \pm m)(2\pm m
    -1)/2}
\end{equation}
(this implies that $S_{-2}=S_2^*$ and $S_{-1}=-S_1^*$).

\section{Correlation functions of polarization patterns and angular power spectra}
\label{observables}

For the sake of completeness, in this appendix we give the relations between the angular power spectra calculated in section \ref{Power-spectra} and observable correlation functions. There is no original material here: these relations are formally the same as for the standard CMB polarization, and we refer the reader to Ref.~\cite{Durrer} for instance, which we follow closely.

It is first convenient to define genuine spin-0 quantities out of the spin-2 Stokes parameters $Q^\Y$ and $U^\Y$. We therefore define 
\begin{eqnarray}
\tilde{E}^\Y({\bf x},{\bf n}) & \equiv & \frac12 \left[  (\spart^*)^2(Q^\Y+i  U^\Y)({\bf x},{\bf n}) +  \spart^2( Q^\Y- i U^\Y)({\bf x},{\bf n})   \right]  \nonumber \msk
&=&\sum_{\ell=2}^{\infty} \sum_{m=-\ell}^\ell \,\sqrt{\frac{(\ell+2)!}{(\ell-2)!}} \,e^\Y_{\ell m}(\gr{x}) \,Y_{\ell m}(\gr{{\hat n}};\gr{{\hat e} })
\end{eqnarray}
and
\begin{eqnarray}
\tilde{B}^\Y({\bf x},{\bf n}) & \equiv & -\frac{i}{2} \left[  (\spart^*)^2( Q^\Y+i  U^\Y)({\bf x},{\bf n}) -  \spart^2( Q^\Y- i U^\Y)({\bf x},{\bf n})   \right] \nonumber \msk
&=&\sum_{\ell=2}^{\infty} \sum_{m=-\ell}^\ell \,\sqrt{\frac{(\ell+2)!}{(\ell-2)!}}\, b^\Y_{\ell m}(\gr{x})\, Y_{\ell m}(\gr{{\hat n}};\gr{{\hat e} })\,,
\end{eqnarray}
where $\spart$ and $\spart^*$ are the spin raising and lowering operators and where $e^\Y_{\ell m}$ and $b^\Y_{\ell m}$ are defined in Eq.~\refeq{def-eb}. Besides being spin-0 quantities, $\tilde{E}^\Y({\bf x},{\bf n})$ and $\tilde{B}^\Y({\bf x},{\bf n}) $ are particularly convenient to adopt because they verify \cite{Durrer}
\begin{equation}
\tilde{E}^\Y= {\rm div} \,{\rm div}\, \Y_{ij}\,, \quad\quad  \tilde{B}^\Y= {\rm rot} \,{\rm rot}\, \Y_{ij}\,.
\end{equation}
Hence $\tilde{E}^\Y$ and $\tilde{B}^\Y$ are locally related to the observable polarization distortion pattern $\Y_{ij}$ and $\tilde{E}^\Y$ measures ``gradient-type'' polarization distortion while $\tilde{B}^\Y$ measures ``curl-type'' polarization distortion. They are directly related to the multipolar coefficients $E_{lm}^\Y$ and $B_{lm}^\Y$ as
\begin{eqnarray}
\tilde{E}^\Y({\bf x},{\bf n}) &\;=\;&  \int \frac{{\rm d}^3 \gr{k}}{(2 \pi)^3} e^{i {\bf k} \cdot {\bf x}} \sum_{\ell=2}^{\infty} \sqrt{\frac{(\ell+2)!}{(\ell-2)!}} \sum_{m=-2}^{2} {E}_{\ell m}^\Y({\bf k})\, Y_{\ell m}({\bf n};{\bf k}) \msk
\tilde{B}^\Y({\bf x},{\bf n}) &\;=\;&  \int \frac{{\rm d}^3 \gr{k}}{(2 \pi)^3} e^{i {\bf k} \cdot {\bf x}} \sum_{\ell=2}^{\infty} \sqrt{\frac{(\ell+2)!}{(\ell-2)!}} \sum_{m=-2}^{2} {B}_{\ell m}^\Y({\bf k})\, Y_{\ell m}({\bf n};{\bf k})
\end{eqnarray}
where $Y_{\ell m}({\bf n};{\bf k})$ indicates that the spherical harmonics are defined with respect to the $z$ axis aligned with the wave vector $\bf{k}$. We find therefore that their angular correlation functions are related to the power-spectra \refeq{def-Cl} by\footnote{There are typos in Eqs.~5.78-5.80 of Ref.~\cite{Durrer}, which are corrected here.}
\begin{eqnarray}
\langle\tilde{E}^\Y({\bf x},{\bf n}) \tilde{E}^\Y({\bf x},{\bf n'})     \rangle &\;=\;& \frac{1}{4 \pi} \,\sum_{\ell=0}^{\infty} \frac{(\ell+2)!}{(\ell-2)!} \,(2\ell+1)\, P_\ell({\bf n} \cdot {\bf n'})\, C_\ell^{E^{\Y}} \msk
\langle \tilde{B}^\Y({\bf x},{\bf n}) \tilde{B}^\Y({\bf x},{\bf n'})     \rangle &\;=\;& \frac{1}{4 \pi} \,\sum_{\ell=0}^{\infty} \frac{(\ell+2)!}{(\ell-2)!} \,(2\ell+1)\, P_\ell({\bf n} \cdot {\bf n'})\, C_\ell^{B^{\Y}}
\end{eqnarray}
where $P_\ell$ denotes the Legendre polynomial of order $\ell$.

\section{Cross-correlation with intensity distortion}\label{Cross}

In this appendix, we calculate the cross-correlation between the $y$-type polarization distortion and the $y$-type intensity distortion. Following the same reasoning as in the main text, we solve the Boltzmann equation for the intensity distortion
\begin{equation}
y' + n^i \partial_i y =
\tau'\left(-y+C^y \right)
\label{y'}
\end{equation}
with the leading-order contribution to the collision term identified as~\cite{Pitrouysky}
\begin{equation}
C^{{y} \,{\rm (L.O.)}}= \frac{1}{3}v_i v^i+\frac{11}{20} v_{\langle i} v_{j\rangle} n^i n^j\,,
\label{Cijy}
\end{equation}
where $\langle ij\rangle$ means the symmetric traceless part.
The corresponding integral solution is
\begin{equation}\label{IntegralFormy}
{y}(\eta_0,k_i,n^{i}) = \int_{\eta_{\re}}^{\eta_0} \dd \eta \,g(\eta)
e^{-\ii k_i
n^{i} r(\eta)} C^{y\, {\rm (L.O.)}}(\eta,k_i,n^{i})\,.
\end{equation}
We then expand in spherical harmonics the contribution of a given Fourier mode, for $y$:
\begin{equation}\label{defylm}
y(\gr{k},\gr{\n}) = \sum_{\ell=2}^{\infty} \sum_{m=-\ell}^\ell
\frac{1}{N_\ell} y_{\ell m}(\gr{k})Y_{\ell m}(\gr{\n})\,,
\end{equation}
and for the collision term:
\begin{equation}
C^{{y}\, {\rm (L.O.)}}(\eta,\gr{k},\gr{\n}) =\sum_{\ell,m}
\frac{1}{N_\ell}C^{y}_{\ell m}(\gr{k})  Y_{\ell m}(\gr{\n})\,.
\label{multipole-Cy}
\end{equation}
Following steps similar to the ones detailed in Appendix \ref{vivj}, we find a monopolar and a quadrupolar contribution:
\begin{equation}
C^{y}_{\ell m} = {\cal K} \left\{\left[
  \delta_\ell^0\delta_m^0 S^y_{00}(\gr{{\hat k}}_1,\gr{{\hat
      k}}_2)+\delta_\ell^2 S^y_{2m}(\gr{{\hat k}}_1,\gr{{\hat k}}_2) \right]
  \F(k_1,\eta)  \F(k_2,\eta)\Phi(k_1)\Phi(k_2)\right\}
  \label{E-Cy}
\end{equation}
where
\begin{equation}
S^y_{00}(\gr{{\hat k}}_1,\gr{{\hat k}}_2)=-\frac{1}{3} \gr{{\hat k}}_1 \cdot \gr{{\hat k}}_2
\label{Smy00}
\end{equation}
and 
\begin{equation}
S^y_{2m}(\gr{{\hat k}}_1,\gr{{\hat k}}_2)=-\frac{11}{\sqrt{6}}S_{m}(\gr{{\hat k}}_1,\gr{{\hat k}}_2)=\frac{11\pi}{45} \sum_{n=-1}^1
\alpha_{n,m}  \left( Y_1^{m-n}(\gr{\hat{k}_1}) Y_1^{n }(\gr{\hat{k}_2}) \right)^*\,.
\label{Smy2m}
\end{equation}
Using the Rayleigh formula, we then obtain
\begin{equation}
\frac{y_{\ell m}(\gr{k})}{2\ell+1}= {\cal K}\left\{\int_{\eta_\re}^{\eta_0}
  \dd \eta g(\eta) \left[j_\ell \delta_m^0 S_{00}^y(\gr{{\hat k}}_1,\gr{{\hat k}}_2)+j_\ell^{(2m)}[k r(\eta)]
S_{2m}^y(\gr{{\hat k}}_1,\gr{{\hat k}}_2) \right] F(k_1,\eta)F(k_2,\eta)   \Phi(k_1) \Phi(k_2)\right\}\,  \label{result-ylm} \\
\end{equation}
where the expressions of the functions $j_l^{(2 m)}$ are given in Appendix \ref{epsilon-beta-lm}. From this the angular power spectra can be obtained by summing all
the Fourier modes contributions, giving
\begin{eqnarray}
(2l+1)^2 C_\ell^{y\,E^\Y} &=& \frac{2}{\pi} \sum_{m=-2}^2 \int {\rm d} k k^2  {\cal Q}^{y\,E^\Y}_{\ell m}(k)\msk
(2l+1)^2 C_\ell^{y} &=& \frac{2}{\pi} \sum_{m=-2}^2 \int {\rm d} k k^2  {\cal Q}^{yy}_{\ell m}(k)
\label{ClyY}
\end{eqnarray}
(the correlation between $y$ and $B^{\Y}$ vanishes by parity), where we defined
\begin{eqnarray}
\langle y_{\ell m}(\gr{k}) E_{\ell m'}^{\Y *}(\gr{k'}) \rangle&=&(2 \pi)^3 \delta^3({\bf k}-{\bf k'})  {\cal Q}^{y\,E^\Y}_{\ell m}(k) \delta_{m m'}\msk
\langle y_{\ell m}(\gr{k}) y_{\ell m'}(\gr{k'}) \rangle&=&(2 \pi)^3 \delta^3({\bf k}-{\bf k'})  {\cal Q}^{yy}_{\ell m}(k) \delta_{m m'}\,.
\end{eqnarray}
From Eqs.~\refeq{result-E}-\refeq{result-ylm}, one easily obtains
\begin{eqnarray}
 {\cal Q}^{yy}_{\ell m}(k)&=&  \frac{2 (2\ell +1)^2}{(2 \pi)^3 } \int
 {\rm d} ^3 \gr{k}_1 P(k_1) P(k_2)   \msk
&&\left|
  \int_{\eta_\re}^{\eta_0} \dd \eta g(\eta) \left\{ S^y_{2m}(\gr{{\hat k}}_1,\gr{{\hat k}}_2) j_\ell^{(2m)}[k r(\eta)]+S^y_{00}(\gr{{\hat k}}_1,\gr{{\hat k}}_2) j_\ell[k r(\eta)] \delta_m^0\right\}
  F(k_1,\eta)F(k_2,\eta) \right|^2\nonumber
 \label{yy}
\end{eqnarray}
and
\begin{eqnarray}
 {\cal Q}^{y\,E^\Y}_{\ell m}(k)&=&  \frac{2 (2\ell +1)^2}{(2 \pi)^3 } \int
 {\rm d} ^3 \gr{k}_1 P(k_1) P(k_2)  \nonumber \msk
&& \times
\left(   \int_{\eta_\re}^{\eta_0} \dd \eta g(\eta) \left\{ S^y_{2m}(\gr{{\hat k}}_1,\gr{{\hat k}}_2) j_\ell^{(2m)}[k r(\eta)]+S^y_{00}(\gr{{\hat k}}_1,\gr{{\hat k}}_2) j_\ell[k r(\eta)] \delta_m^0 \right\}
  F(k_1,\eta)F(k_2,\eta) \right) \nonumber\msk
&&\times  \left( S^\star_{m}(\gr{{\hat k}}_1,\gr{{\hat k}}_2) \int_{\eta_\re}^{\eta_0} \dd \eta g(\eta) \epsilon_\ell^{(m)}[k r(\eta)] F(k_1,\eta)F(k_2,\eta) \right) \,,
 \label{yE-Y}
\end{eqnarray}
with the same notations as in \S~\ref{Limber}. Following the same arguments as there, we find that the auto-correlation of intensity distortion reads, within the Limber approximation:
\begin{eqnarray}\label{EqLimberyy}
C_{\ell\, {\rm Limber}}^{y} &=& \frac{2}{(2 \pi)^2} \int_0^{r_\re} \frac{{\rm d} r }{r^2}\, k_1^2 {\rm d} k_1 \sin
\theta_{\gr{k}_1} \dd \theta_{\gr{k}_1} P(k_1) P(k_2) \left(  g(\eta )F(k_1,\eta) F(k_2,\eta ) \right)^2 \nonumber \msk
&& \qquad \qquad  \times \left(\left|S^y_{00}(\gr{{\hat k}}_1,\gr{{\hat k}}_2)+ \frac{1}{2}S^y_{20}(\gr{{\hat k}}_1,\gr{{\hat k}}_2)\right|^2+\frac{3}{4}\left|S^y_{22}(\gr{{\hat k}}_1,\gr{{\hat k}}_2)\right|^2  \right)\,.
\end{eqnarray}
Similarly, the cross correlation between intensity distortion and
polarization distortion is, in the Limber approximation:
\begin{eqnarray}
C_{\ell\, {\rm Limber}}^{y\,E^\Y} &=& \frac{2}{(2 \pi)^2} \int_0^{r_\re} \frac{{\rm d} r }{r^2}\, k_1^2 {\rm d} k_1 \sin
\theta_{\gr{k}_1} \dd \theta_{\gr{k}_1} P(k_1) P(k_2) \left(  g(\eta )F(k_1,\eta) F(k_2,\eta ) \right)^2 \msk
&& \quad   \times \left\{\sqrt{\frac{3}{8}}\left[S^y_{00}(\gr{{\hat
          k}}_1,\gr{{\hat k}}_2) +\frac{1}{2}S^y_{20}(\gr{{\hat
          k}}_1,\gr{{\hat
          k}}_2)\right]S^\star_{0}(\gr{{\hat
        k}}_1,\gr{{\hat k}}_2)-\frac{1}{4}\sqrt{\frac{3}{2}}S^y_{22}(\gr{{\hat
        k}}_1,\gr{{\hat k}}_2) S^\star_{2}(\gr{{\hat k}}_1,\gr{{\hat k}}_2)  \right\} \,. \nonumber
\end{eqnarray}
It is displayed on the right panel of Fig.~\ref{fig:SpectralFunctions}.

\section{Expressions of special functions}
\label{epsilon-beta-lm}

We first give the expressions of $\epsilon_\ell^{(m)}$ and $\beta_\ell^{(m)}$ entering into the results Eqs.~\refeq{result-E}-\refeq{result-B} for the spectral distortions $E$- and $B$-modes:
\begin{eqnarray}
\epsilon^{(0)}_\ell(x) & = &\sqrt{{3 \over 8 }
	{(\ell+2)! \over (\ell-2)!}} {j_\ell(x) \over x^2} \, ,
	\vertsp\nonumber\msk
\epsilon^\ve_\ell(x) &=& {1 \over 2} \sqrt{(\ell-1)(\ell+2)}
        \left[ {j_\ell(x) \over  x^2} + {j_\ell'(x) \over  x}
        \right] \, , \nonumber\msk
\epsilon^\te_\ell(x) &=& {1 \over 4} \left[ -j_\ell(x)
        + j_\ell''(x) + 2{j_\ell(x) \over x^2} +
        4{j_\ell'(x) \over x} \right]  \, 
\label{eqn:epsilon}
\end{eqnarray}
and
\begin{eqnarray}
\beta^{(0)}_\ell(x) &=& 0\, , \nonumber\msk
\beta^{(1)}_\ell(x) &=& {1 \over 2} \sqrt{(\ell-1)(\ell+2)}
        {j_\ell(x) \over x} \, , \nonumber\msk
\beta^{(2)}_\ell(x) &=& {1 \over 2} \left[ j_\ell'(x)
        + 2 {j_\ell(x) \over x} \right] \, ,
\label{eqn:beta}
\end{eqnarray}
where $j_\ell(x)$ denotes the spherical Bessel function of order $\ell$, and with $\epsilon_\ell^{(-1)}=\epsilon_\ell^{(1)}$, $\epsilon_\ell^{(-2)}=\epsilon_\ell^{(2)}$, $\beta_\ell^{(-1)}=- \beta_\ell^{(1)}$ and $\beta_\ell^{(-2)}=- \beta_\ell^{(2)}$.\\

We then give the expressions of $j_\ell^{(2 m)}$ entering into the result Eq.~\refeq{result-ylm} for the intensity spectral distortion:
\begin{eqnarray}
j_\ell^{(20)}(x)&=& {1 \over 2} [3 j_\ell''(x) + j_\ell(x) ]\,, \msk
j_\ell^{(21)}(x) &=& \displaystyle{ 
	\sqrt{3\ell(\ell+1) \over 2}\, \left({j_\ell(x) \over x}
	\right)' }    
	\vphantom{\Bigg[}  \,, \msk			  
			  j_\ell^{(22)}(x) &=& \displaystyle{ 
	\sqrt{{3 \over 8} {(\ell+2)! \over (\ell-2)!}} \,
		{j_\ell(x) \over x^2} } 
	\vphantom{\Bigg[} \,,
\label{eqn:jdef}
\end{eqnarray}
with $j_\ell^{(2\,-2)}=j_\ell^{(2 2)}$ and $j_\ell^{(2\,-1)}=j_\ell^{(2 1)}$.


\end{document}